\documentclass[twocolumn]{aastex631}

\usepackage{booktabs,chemformula}

\graphicspath{{./}}

\received{}
\revised{}
\accepted{}
\submitjournal{ApJ Letters}

\shorttitle{Galaxy cluster FRBs}
\shortauthors{Connor et al.}

\begin{document}

\title{Deep Synoptic Array science: Two fast radio burst sources in massive galaxy clusters}

\correspondingauthor{Liam Connor}
\email{liam.dean.connor@gmail.com}

\author{Liam Connor}
\affiliation{Cahill Center for Astronomy and Astrophysics, MC 249-17 California Institute of Technology, Pasadena CA 91125, USA.}

\author{Vikram Ravi}
\affiliation{Cahill Center for Astronomy and Astrophysics, MC 249-17 California Institute of Technology, Pasadena CA 91125, USA.}
\affiliation{Owens Valley Radio Observatory, California Institute of Technology, Big Pine CA 93513, USA.}

\author{Morgan Catha}
\affiliation{Owens Valley Radio Observatory, California Institute of Technology, Big Pine CA 93513, USA.}

\author{Ge Chen}
\affiliation{Cahill Center for Astronomy and Astrophysics, MC 249-17 California Institute of Technology, Pasadena CA 91125, USA.}

\author{Jakob T. Faber}
\affiliation{Cahill Center for Astronomy and Astrophysics, MC 249-17 California Institute of Technology, Pasadena CA 91125, USA.}

\author{James W. Lamb}
\affiliation{Owens Valley Radio Observatory, California Institute of Technology, Big Pine CA 93513, USA.}

\author{Gregg Hallinan}
\affiliation{Cahill Center for Astronomy and Astrophysics, MC 249-17 California Institute of Technology, Pasadena CA 91125, USA.}
\affiliation{Owens Valley Radio Observatory, California Institute of Technology, Big Pine CA 93513, USA.}

\author{Charlie Harnach}
\affiliation{Owens Valley Radio Observatory, California Institute of Technology, Big Pine CA 93513, USA.}

\author{Greg Hellbourg}
\affiliation{Cahill Center for Astronomy and Astrophysics, MC 249-17 California Institute of Technology, Pasadena CA 91125, USA.}
\affiliation{Owens Valley Radio Observatory, California Institute of Technology, Big Pine CA 93513, USA.}

\author{Rick Hobbs}
\affiliation{Owens Valley Radio Observatory, California Institute of Technology, Big Pine CA 93513, USA.}

\author{David Hodge}
\affiliation{Cahill Center for Astronomy and Astrophysics, MC 249-17 California Institute of Technology, Pasadena CA 91125, USA.}

\author{Mark Hodges}
\affiliation{Owens Valley Radio Observatory, California Institute of Technology, Big Pine CA 93513, USA.}

\author{Casey Law}
\affiliation{Cahill Center for Astronomy and Astrophysics, MC 249-17 California Institute of Technology, Pasadena CA 91125, USA.}
\affiliation{Owens Valley Radio Observatory, California Institute of Technology, Big Pine CA 93513, USA.}

\author{Paul Rasmussen}
\affiliation{Owens Valley Radio Observatory, California Institute of Technology, Big Pine CA 93513, USA.}

\author{Jack Sayers}
\affiliation{Cahill Center for Astronomy and Astrophysics, MC 249-17 California Institute of Technology, Pasadena CA 91125, USA.}

\author{Kritti Sharma}
\affiliation{Cahill Center for Astronomy and Astrophysics, MC 249-17 California Institute of Technology, Pasadena CA 91125, USA.}

\author{Myles B. Sherman}
\affiliation{Cahill Center for Astronomy and Astrophysics, MC 249-17 California Institute of Technology, Pasadena CA 91125, USA.}

\author{Jun Shi}
\affiliation{Cahill Center for Astronomy and Astrophysics, MC 249-17 California Institute of Technology, Pasadena CA 91125, USA.}

\author{Dana Simard}
\affiliation{Cahill Center for Astronomy and Astrophysics, MC 249-17 California Institute of Technology, Pasadena CA 91125, USA.}

\author{Jean Somalwar}
\affiliation{Cahill Center for Astronomy and Astrophysics, MC 249-17 California Institute of Technology, Pasadena CA 91125, USA.}

\author{Reynier Squillace}
\affiliation{Cahill Center for Astronomy and Astrophysics, MC 249-17 California Institute of Technology, Pasadena CA 91125, USA.}

\author{Sander Weinreb}
\affiliation{Cahill Center for Astronomy and Astrophysics, MC 249-17 California Institute of Technology, Pasadena CA 91125, USA.}

\author{David P. Woody}
\affiliation{Owens Valley Radio Observatory, California Institute of Technology, Big Pine CA 93513, USA.}

\author{Nitika Yadlapalli}
\affiliation{Cahill Center for Astronomy and Astrophysics, MC 249-17 California Institute of Technology, Pasadena CA 91125, USA.}

\collaboration{200}{(The Deep Synoptic Array team)}

\begin{abstract}
The hot gas that constitutes the 
intracluster medium (ICM) has been studied   
at X-ray and millimeter/sub-millimeter wavelengths (Sunyaev–Zeld'ovich effect) for decades. Fast radio bursts (FRBs) offer an additional 
method of directly measuring the ICM and gas surrounding clusters, via observables such as dispersion 
measure (DM) and Faraday rotation measure (RM). We report the discovery of two FRB sources detected with the Deep Synoptic Array (DSA-110)
whose host galaxies belong to massive galaxy clusters. 
In both cases, the FRBs exhibit excess extragalactic 
DM, some of which likely originates in the ICM of their respective 
clusters. FRB\,20220914A resides in the galaxy cluster 
Abell\,2310 at $z=0.1125$ with a projected offset from the 
cluster center of 520$\pm$50\,kpc. The host of 
a second source, FRB\,20220509G, is an elliptical galaxy 
at $z=0.0894$ that belongs to the galaxy cluster Abell\,2311
at projected offset of 870$\pm$50\,kpc. 
These sources represent the first time an FRB has been localized to a galaxy cluster. We combine 
our FRB data with archival X-ray, SZ, and optical observations
of these clusters in order to infer properties of the ICM, including a measurement of gas temperature from DM and $y_{SZ}$ of 0.8--3.9\,keV.
We then compare our results to massive cluster halos 
from the IllustrisTNG simulation. Finally, 
we describe how large samples of localized FRBs from 
future surveys will constrain the ICM, particularly beyond the virial radius of clusters.
\end{abstract}


\section{Introduction} \label{sec:intro}

Galaxy clusters are massive ($10^{14}-10^{15}\,M_\odot$) gravitationally bound 
objects comprised of hundreds to thousands of 
galaxies. Galaxies make up 
just a few percent of the total cluster mass.
The dominant component of baryons by mass and volume is in the hot intracluster medium (ICM) made of diffuse gas 
with $n_e\approx10^{-3}$\,cm$^{-3}$ and $T_e\approx10^7-10^8$\,K. Within a sufficiently 
large volume (e.g., the virial radius), the 
ratio of dark matter to baryons approximately matches the 
Universal value \citep{Eckert2019}, which is not true for smaller halos where feedback is expected 
to expel gas beyond the virialized dark matter halos \citep{Tumlinson17}.

The ICM has been studied in great detail at X-ray 
wavelengths for the past fifty years \citep{cluster-xray}. Thermal bremsstrahlung emission 
from gas heated to 2--15\,keV 
is observed out to the virial radii of clusters (1--2\,Mpc), tracing the 
radial distribution of the ICM gas \citep{2019A&A...621A..41G}. The ICM 
reaches such high temperatures by adiabatic compression 
and shock heating as the gas reaches 
hydrostatic equilibrium within the potential well \citep{cluster-reviewx}. 
Free-free emission scales with emissivity, and the 
specific X-ray luminosity can be written as an integral of 
plasma density squared,

\begin{equation}
    L_X \approx \int\,n^2_e\,\Lambda(T_e)\,\textup{d}V,
\end{equation}

\noindent where $n_e$ is the number density of 
free electrons and $\Lambda(T_e)$ characterizes the temperature 
dependence of X-ray emission. The total mass can then be inferred 
from the cluster's X-ray luminosity \citep{Pratt}. X-ray spectroscopy can also be used to measure the gas temperature \citep{x-ray-temp} and its velocity structure \citep{x-ray-velocity}.

The hot plasma in the ICM is also observable at millimeter/sub-millimeter wavelengths through the 
thermal Sunyaev-Zel’dovich (SZ) effect, whereby photons in the cosmic microwave background (CMB) are inverse Compton scattered by thermal electrons \citep{tsz}. This leads
to spectral distortions in the CMB at the level of $10^{-4}$--$10^{-5}$. SZ observations of the ICM 
are parameterized by the Compton 
Y-parameter, which is given by the integral 
of electron thermal pressure along the line of 
sight,

\begin{equation}
    y_{SZ} = \frac{k_B\,\sigma_T}{m_e\,c^2}\int n_e\,T_e\,\textup{d}l,
    \label{eq:sz}
\end{equation}

\noindent where $k_B$ is Boltzmann's
constant, $\sigma_T$ is the Thomson cross-section, 
$m_e$ is electron mass, $c$ is the speed of light. 

Fast radio bursts (FRBs) offer an additional probe of 
the ICM of massive galaxy clusters.
FRBs are short-duration (10$^{-2}$--10$^2$\,ms) radio transients that have been detected over 
cosmological distances ($z\lesssim1.5$) \citep{petroffreview, cordes-review}. To date, $\mathcal{O}(10^3)$ sources have been discovered 
\citep{chime-catalog1, frbcat}
but only two-dozen have been localized with 
sufficient angular precision to identify a 
host galaxy \citep{chatterjee17, ravi19frb, heintz2020, bhandari2020}. 
Propagation effects imparted on the radio pulse 
encode information about the plasma through which 
the burst travelled. Dispersion 
measure (DM) is an integral of line-of-sight 
electron density and is given by,

\begin{equation}
    \textup{DM} = \int\,n_e\,\textup{d}l.
\label{eq:DM}
\end{equation}

Unlike with thermal X-ray emission or SZ observations 
of the ICM, the FRB DM is detectably impacted by  \textit{all}  
plasma between us and the source, not just $10^{6-8}$\,K 
gas. This is both a blessing and a curse: FRB DMs 
will probe plasma in the host galaxy, 
the circumgalactic medium (CGM), the ICM of intervening or host clusters, the intergalactic medium (IGM), and the Milky Way's interstellar 
medium (ISM). For individual FRB sightlines, these terms 
must be explicitly modelled. 
An FRB that is impacted by a galaxy cluster 
will have the following ``DM budget'',

\begin{equation}
    \rm{DM}_{obs} = \textup{DM}_{MW} + \textup{DM}_{IGM} + \frac{\textup{DM}_{ICM}}{1+z_{c}} + \frac{\textup{DM}_{host}}{1+z_{frb}},
\label{eq:dmbudget}
\end{equation}

\noindent where $z_{c}$ and $z_{frb}$ are the cluster and FRB redshifts, respectively. The observed DM is likely dominated by the IGM for most FRBs. Still, it is important to be specific about what is meant by 
the IGM. In \citet{macquart-dm}, the relationship 
between DM$_{\rm cosmic}$ and source redshift includes all extragalactic gas between the Milky Way and the host galaxy. DM$_{\rm cosmic}$ therefore includes the CGM, intragroup medium (IGrM), ICM, and IGM. All of those terms correlate with source distance because the optical depth of halos 
increases with redshift. In this work we take the IGM to be ionized gas that resides outside of virialized dark matter halos \citep{mcquinn-review}.

For sources that intersect or are 
embedded in galaxy clusters, the largest 
DM contribution may be from the ICM \citep{prochaska2019b}. For example, a
cluster with $\langle n_e \rangle=10^{-3}$\,cm$^{-3}$ and 
$R_{200}=1$\,Mpc will lead to DM$_{\rm ICM}\sim10^3$\,pc\,cm$^{-3}$ for low impact parameters. 
This value is more than double the typical contribution from the 
IGM for a source at $z=0.5$ and 70$\%$ larger 
than the mean DM$_{obs}$ of CHIME/FRB 
sources \citep{macquart-dm, chime-catalog1}. The polarization properties of 
FRBs can help measure the magnetic field of the 
ICM. Faraday rotation measure (RM) is an integral 
of the free electrons, weighted by the 
line-of-sight magnetic field strength,

\begin{equation}
    \textup{RM} =  \frac{e^3}{2\pi\,m^2_e\,c^4} \int\,B_{\parallel}\,n_e(l)\,\textup{d}l,
\end{equation}

\noindent where $e$ is the charge of an electron. While other radio objects can provide 
cluster RMs \citep{RMweer}, FRBs are unique amongst extragalactic sources 
in their ability to measure both RM and DM. 
Therefore, if one can isolate the components of 
RM and DM that are due to the ICM, the mean line-of-sight 
magnetic field strength in the cluster can be determined by their ratio, 

\begin{equation}
    \langle B_{\parallel, \rm ICM} \rangle = 1.23\,\mu \textup{G}\,\,\frac{\textup{RM}_{\rm ICM}}{\textup{DM}_{\rm ICM}},
\label{eq:Bperp}
\end{equation}

\noindent where RM and DM are given in the 
standard units of rad\,m$^{-2}$ and pc\,cm$^{-3}$, respectively. The application of FRBs to the ICM 
has been considered in previous works,
including the detection of an FRB in the 
direction of the Virgo cluster by ASKAP \citep{Devansh2019}. In that case, 
the localization precision was insufficient to 
determine a host galaxy, so it was unclear if the source 
was embedded in, or behind, the cluster. Another unlocalized source, FRB\,20190116A \citep{anderson-chime} is likely within a $\sim$\,5\,Mpc of the Coma cluster in projection and may be dispersed by filamentary structure extending from the cluster \citep{hallinan-coma}. \citet{whimfrb2017} 
investigated the application of FRBs to the warm hot intergalactic medium (WHIM) at the outskirts of 
galaxy clusters, where X-ray observations are difficult 
due to the $n_e^2$ dependence of emissivity.
Nonetheless, most of the effort in studying halo gas 
with FRBs has been devoted to the circumgalactic medium \citep{prochaska2019b, ravi2019, kglee}.
That is because the CGM is more difficult to detect than the ICM 
using traditional means
(X-ray and SZ observations are much less 
sensitive to the gas in galaxy-scale halos), and because 
of the CGM's significance in galaxy formation and 
feedback processes. \citet{connorravi} presented the first statistical evidence for the impact 
of halo gas on FRBs and demonstrated the importance of considering galaxy groups 
on FRB DM budgets. As we demonstrate, current and future 
FRB surveys will find a significant number of sources that 
are impacted by galaxy clusters, and understanding 
their effects and prevalence are important for CGM 
studies, in addition to studying the ICM itself. 

In this work we make the first unambiguous detection of  
FRB sources that belong to a galaxy cluster. 
Both sources were discovered by the
real-time FRB survey on the Deep Synoptic Array (DSA-110)\footnote{\url{https://deepsynoptic.org}}. 
The DSA-110 is a radio interferometer at the Owens Valley Radio Observatory (OVRO) that was built to localize FRBs with $\sim$\,arcsecond
precision.
A detailed account of the instrument will be presented in Ravi et al. (in prep.). During the observations presented in this work, the DSA-110 was operating 
in its commissioning phase with 63 antennas collecting data, as described in \citet{rcc+22}.

In Section~\ref{sec:observations}
we describe the radio properties of two FRBs using  
the DSA-110.
We describe the two massive clusters of which the 
FRB host galaxies are members. A companion paper details the interferometric localizations of FRB\,20220509G and 
FRB\,20220914A as well as their unusual host galaxies ~(Sharma et al., in prep.). 
In Section~\ref{sec:dmbudget} we 
model the DM contribution from the ICM for both 
sources, combining our radio data with archival X-ray, 
SZ, and optical data to infer properties about the 
clusters' respective ICM. Finally, 
we compare our results with clusters found in the 
IllustrisTNG simulation.

\begin{table*}[t]
\centering
\begin{tabular}{@{}cccccccccc@{}}
\toprule
\textbf{Source}        & \textbf{RA (deg)} & \textbf{Dec (deg)} & $\mathbf{z_{gal}}$ & \textbf{DM$_{obs}$} & \textbf{DM}$_{MW}$ & \textbf{RM}$_{obs}$ & $\mathbf{\tau}$ \textbf{(ms)}    & \textbf{$\mathcal{F}$ (Jy ms)} & \textbf{$b_\perp$ (kpc)} \\ \cline{1-10}
FRB\,20220914A & 282.0568 & 73.3369 & 0.1139  & 631.29     & 47/55      & N/A       & $<0.080$       & 2.2      & 520$\pm$50      \\
FRB\,20220509G & 282.6700 & 70.2438 & 0.0894 & 269.53      & 46/55      & $-110$(1)    & 0.08$\pm$0.02 & 5.5 & 870$\pm$50 \\
\bottomrule
\end{tabular}
\caption{The radio properties of two galaxy cluster 
FRBs. DM and RM are reported in pc\,cm$^{-3}$ and 
rad\,m$^{-2}$, respectively. The scattering timescale, 
$\tau$, is referenced to 1.4\,GHz. The two Galactic ISM DM estimates 
listed are YMW16 and NE2001. RA and Dec are in epoch J2000.}
\label{tab:frb}
\end{table*}

\begin{table*}[t]
\centering
\begin{tabular}{@{}ccccccccc@{}}
\toprule
\textbf{Cluster} & \textbf{RA (deg)} & \textbf{Dec (deg)}                                    & \textbf{Richness} & $\mathbf{z_{clust}}$ & \textbf{M$_{500}$} ($M_\odot$)                    & \textbf{R$_{500}$ (kpc)} & \textbf{L$_X$ (erg/s)} & \textbf{Y$_{5R500}$ (arcmin$^2$)} \\ \cline{1-9}
Abell\,2310       &     281.81708              &   73.3472                                                    & 124               & 0.1125            &                 1.7$\times$10$^{14}$                         &                  810     &         $8.3\times10^{43}$        &  $0.85\pm0.35 \times 10^{-3}$ 
\\
Abell\,2311       & 282.497           & 70.376 & 190               & 0.0899(1)         & 1.6$\times10^{14}$ & 800                      & 4.7$\times10^{43}$ &    $0.81\pm0.36 \times 10^{-3}$          \\
\bottomrule
\end{tabular}
\caption{The properties of two galaxy clusters that host the FRBs presented in this paper. $L_X$ is the X-ray luminosity at 0.1--2.4\,keV
within $R_{500}$.}
\label{tab:clusters}
\end{table*}

\section{Cluster FRBs}
\label{sec:observations}
The radio properties of FRB\,20220914A and 
FRB\,20220509G are listed in Table~\ref{tab:frb}. The 
observed parameters of their respective host galaxy clusters 
are given in Table~\ref{tab:clusters}. The two FRBs presented here are from distinct galaxy clusters, but are relatively nearby on the sky---separated by just 3.1 degrees. DSA-110 is a transit 
instrument that was parked at Declination 
$\sim$\,$+$70 for most of the 2022 commissioning period. For this reason, most of our initial sample of sources fall on a ring in RA and several pairs are within a few degrees of each other. Below 
we described observations of the FRBs and 
their host galaxies, as well as archival data obtained for the galaxy clusters.

\begin{figure}[h]
    \centering
    \includegraphics[width=0.3\textwidth]{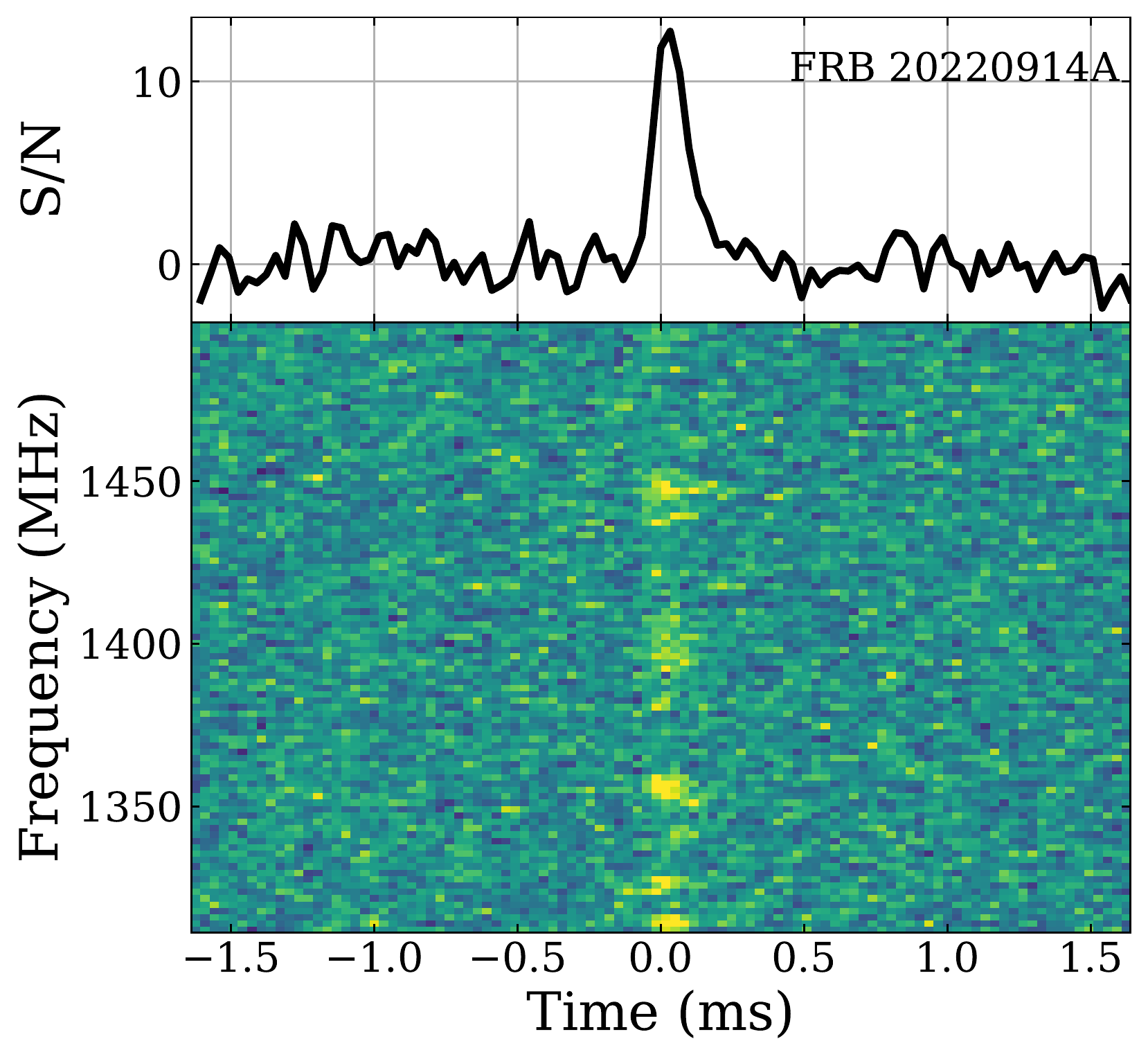}\\
    \includegraphics[width=0.3\textwidth]{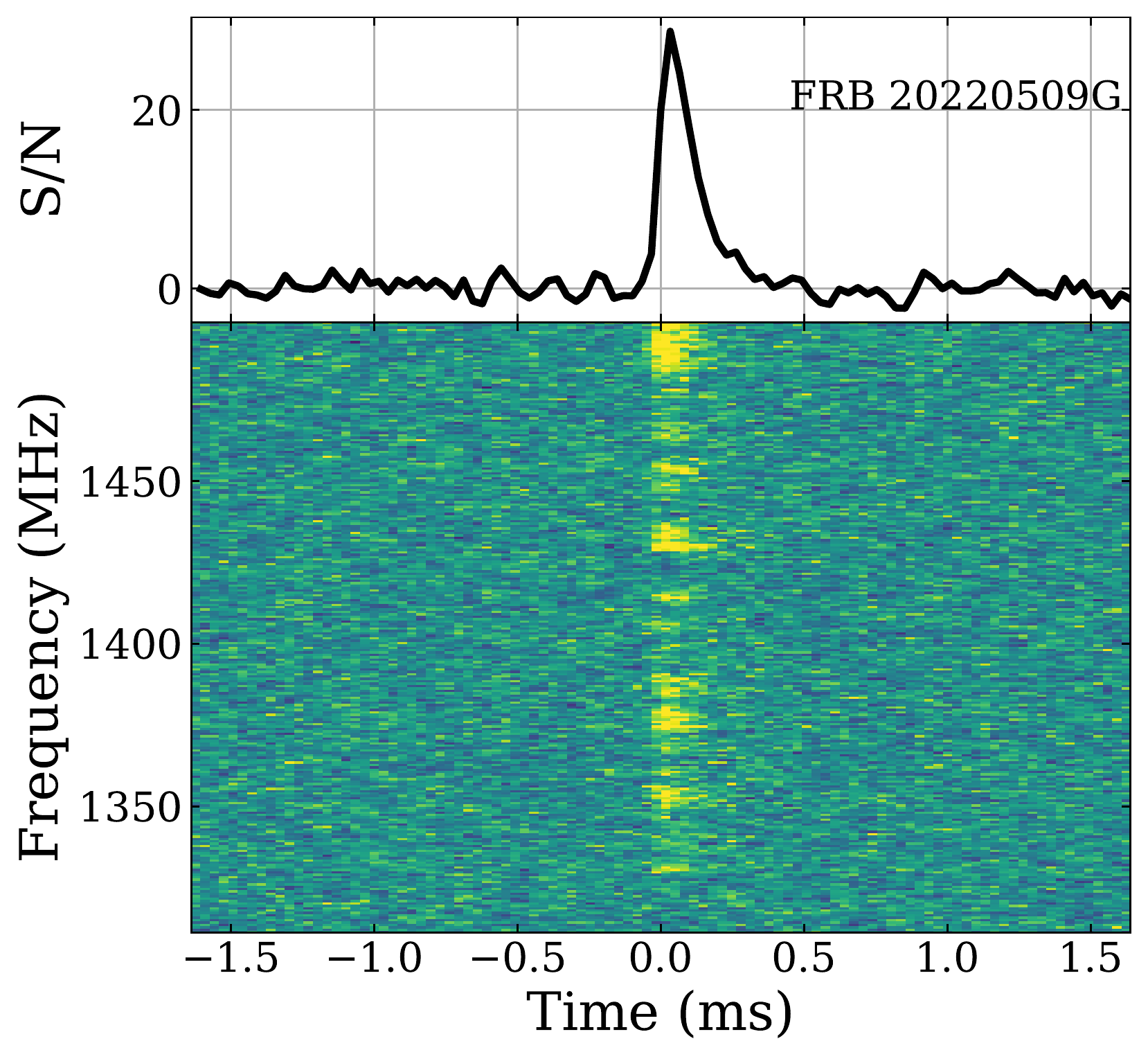}
    \caption{Dedispersed total-intensity dynamic spectra for two FRBs 
    detected by DSA-110. Both 
    FRBs reside in galaxies that 
    belong to massive galaxy clusters.}
    \label{fig:bursts}
\end{figure}

\subsection{FRB\,20220914A}

FRB\,20220914A was detected at 
MJD 59836.1459660. Follow-up analysis  
found an optimal DM of 631.3\,pc\,cm$^{-3}$ and a 
pulse full-width at half maximum (FWHM) of 140\,$\mu$s.
With a S/N of 
13.4 and system-equivalent flux density (SEFD) 
of 140\,Jy, the fluence of the burst was 
roughly 2.2\,Jy\,ms. 
Offline interferometric localization 
found the position of the 
FRB to be R.A. J2000, decl. J2000 = 18h48m13.63s, $+$73d20m12.89s, the centroid of which is 
1.6'' from a galaxy with spectroscopic
redshift $z=0.1139$. 
The expected DM from the Milky 
Way at these Galactic coordinates is 
roughly 50\,pc\,cm$^{-3}$ from NE2001 \citep{ne2001} and 
47\,pc\,cm$^{-3}$ from YMW16 \citep{ymw16}. The IGM 
is expected to contribute 50--120\,pc\,cm$^{-3}$ at this redshift,
suggesting that FRB\,20220914A has significant extragalactic 
DM excess. 

We find no evidence of scattering, placing 
an upper-limit of $\tau<80$\,$\mu$s at 1.4\,GHz. Polarimetric analysis found no detectable 
polarization and no RM was determined after searching a range of $-10^6$ to $+10^6$\,rad\,m$^{-2}$.
The upper-limits on polarization fraction were 
$L/I<15\%$ and $V/I<20\%$.

\subsubsection{ABELL\,2310 galaxy cluster}
The position of FRB\,20220914A was cross-matched with 
both the DESI Legacy Imaging Surveys Data Release 9 (DR9) galaxy catalog \citep{desi} and DR9 cluster/group catalog \citep{desicluster}. 
We find that the host galaxy of FRB\,20220914A 
is a member of the massive galaxy cluster 
Abell\,2310 \citep{Abell}. According to the DESI DR9 group/cluster catalog, this cluster has 
richness 124 (i.e. number of member galaxies), 
$M_{180}=2.5\times10^{14}$\,$M_\odot$, with a 
brightest cluster galaxy (BCG) at spectroscopic redshift $z=0.1125$. 
The cluster is also in the Meta-Catalog of X-Ray Detected Clusters of Galaxies (MCXC) with source name J1847.2+7320 \citep{mcxc2011}. The X-ray surface brightness centroid is at 18h47m16.s 73d20m50s \citep{mcxc2011}. The FRB offset from 
this position is 4.22' suggesting a projected physical impact parameter of 
520\,kpc. The X-ray luminosity within $R_{500}$ is 8.31$\times10^{43}$\,erg\,s$^{-1}$, which gives
$M_{500}=1.69\times10^{14}$\,$M_\odot$ using an 
empirical $L_X-M$ relation \citep{pratt2009, mcxc2011}. 

The cluster has also been detected via the 
thermal SZ effect and is listed in the 
Planck SZ2 cluster catalog as PSZ2\,G104.29+26.17 \citep{PSZ2}. 
The SZ-derived mass based on a hydrostatic mass calibration is $M_{500} = 1.94\pm0.28\times10^{14}\,M_\odot$ \citep{planck14},
and its SZ centroid has a physical offset from FRB\,20220914A 
of $\sim$\,380\,kpc. We have also analyzed the Planck MILCA $y$-map from \citet{Planck2016_ymap} using the techniques described in \citet{sayers16} to obtain both a two-dimensional projected model of the cluster $y(\theta)$, along with the total integrated SZ signal $Y_{5R500}$. In brief, we assume the cluster follows the profile shape given by \citet{Arnaud2010}, with the scale radius set by the X-ray value of $R_{500}$ and centered on the X-ray centroid. We then determine the best-fit normalization of this model, with the uncertainty on this normalization estimated from fits to 100 random realizations of the Planck noise. The result is $Y_{5R500} = 0.85 \pm 0.35$~arcmin$^2$.

The point-spread function (PSF) of Planck is large, so we take the offset between FRB\,20220914A and Abell\,2310 
to be the separation between the X-ray centroid and the FRB's host galaxy. 
This gives a projected physical offset of 520$\pm$50\,kpc. 
In Figure~\ref{fig:cluster-elektra} we show the filtered X-ray data with a contour for the SZ emission region, as well as a zoom in on the optical host galaxy image. The X-ray 
intensity data is taken from the public archive of 
X-ray clusters\footnote{\url{https://github.com/wwxu/rxgcc.github.io}} \citep{weiwei-rosat}.

\begin{figure}[h]
    \centering
    \includegraphics[width=0.475\textwidth]{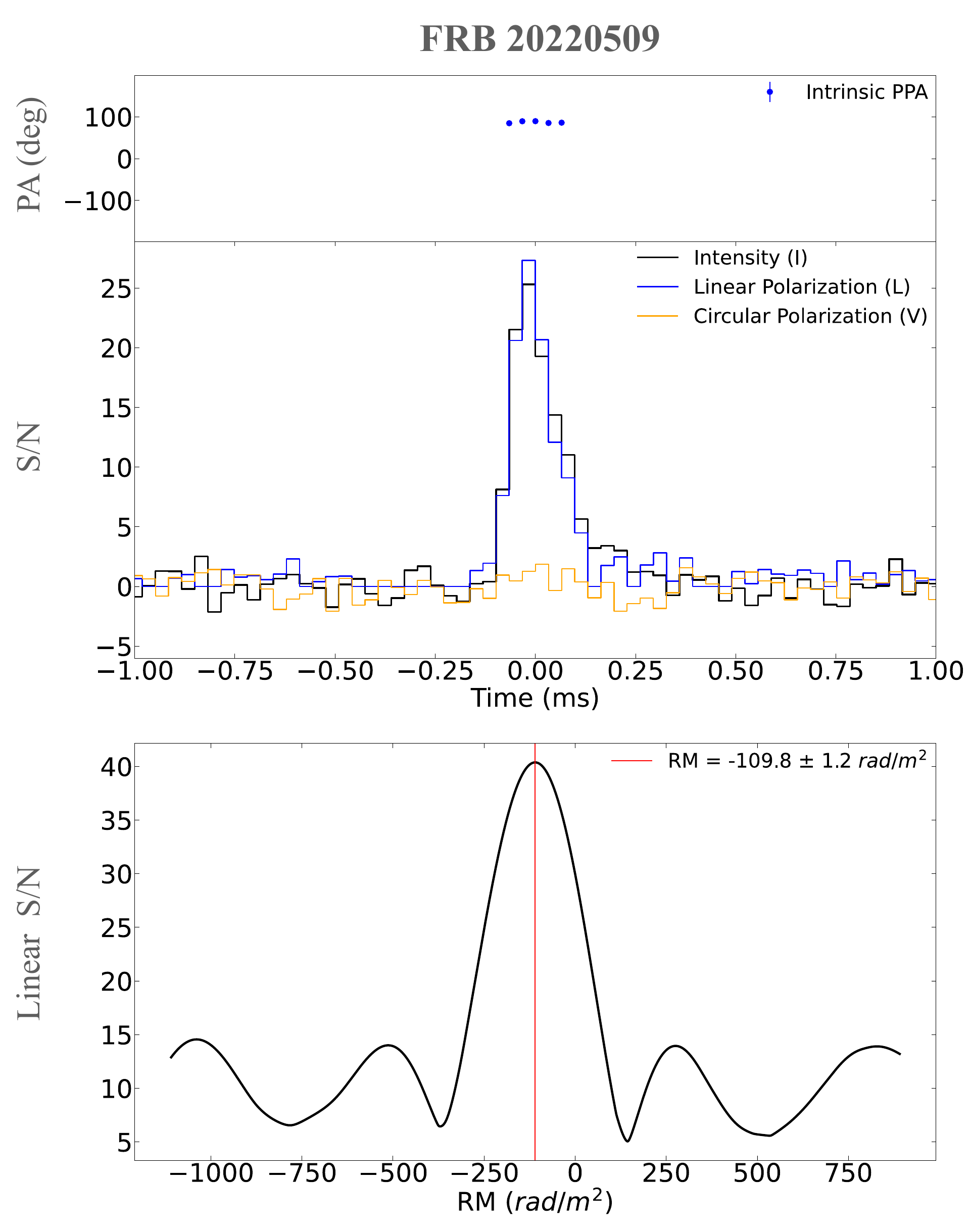}
    \caption{The polarization properties of 
    FRB\,20220509G. The top panel shows a 
    flat PA across the pulse and the middle 
    panel shows the linearly polarized pulse profile.
    The bottom panel is the Faraday spectrum with a 
    peak at RM=$-111.5$\,rad\,m$^{-2}$.}
    \label{fig:jackiepol}
\end{figure}

The redshift difference between the host galaxy of FRB\,20220914A and the brightest cluster galaxy (BCG) is $\Delta z=-0.00129$, 
indicating the host has a recession velocity of roughly 390\,km\,s$^{-1}$. We do not know the radial position of the FRB within the cluster, but we have good reason 
to believe it is not near the front. We take the radial 
position to be $l_{LOS}$, which is zero at the distance of the cluster center and negative behind the cluster. The radial distribution of galaxies in clusters is well described by an NFW profile with concentration parameter c=2.6 \citep{galdist}. With a virial radius of 800\,kpc, the probability distribution of galaxy radius peaks around 300\,kpc and declines towards larger radii. Therefore, for an impact parameter beyond 300\,kpc, the most likely $l_{LOS}$ is zero. More convincingly, as we show in Section~\ref{sec:dmbudget}, the FRB has significant excess dispersion that likely comes from the ICM. The magnitude of this extra DM 
cannot easily be explained by the IGM or the host galaxy, so $l_{LOS}$ is likely smaller than several hundred kpc.

\subsubsection{The host galaxy of FRB\,20220914A}

The FRB 20220914 has been localized to a typical late-type spiral galaxy at R.A. (J2000) = 18:48:13.9580 and declination (J2000) = +73:20:10.703 ~(Sharma et al., in prep.). The spectroscopic redshift of the galaxy, as measured from optical spectra acquired with Keck-I/LRIS, is $0.1139\pm0.0001$. A detailed spectral energy distribution analysis revealed a stellar mass of $\log M_* (M_\odot) = 9.99^{+0.09}_{-0.09}$. The constrained star formation history indicates significant recent star formation with several starbursts over the last 3.5~Gyr, thus reflecting a wide probability distribution for the age of its progenitor.


\subsection{FRB\,20220509G}
The fast radio burst source FRB\,20220509G 
had an arrival time of 
59708.4944991 at reference frequency 
of 1500\,MHz. Its optimal DM was
269.53\,pc\,cm$^{-3}$ and had a fluence of 5.5\,Jy\,ms. 
Offline interferometric localization found that the 
source is at R.A. J2000, decl. J2000 = 18h50m40.8s, $+$70d14m37.8, with a 90$\%$ error 
ellipse with axes 4.7 and 3.2 in R.A. and decl. This 
is within 6'' of a massive galaxy with spectroscopic 
redshift $z=0.0894$. The Galactic ISM DM estimates 
are 55\,pc\,cm$^{-3}$ from NE2001 and 46\,pc\,cm$^{-3}$ from YMW16. 
The expected IGM contribution at this redshift is roughly 75\,pc\,cm$^{-3}$, but with significant uncertainty. We analyze 
the source's DM budget in Section~\ref{sec:dmbudget}.
\begin{figure}
    \centering
    \includegraphics[width=0.5
    \textwidth]{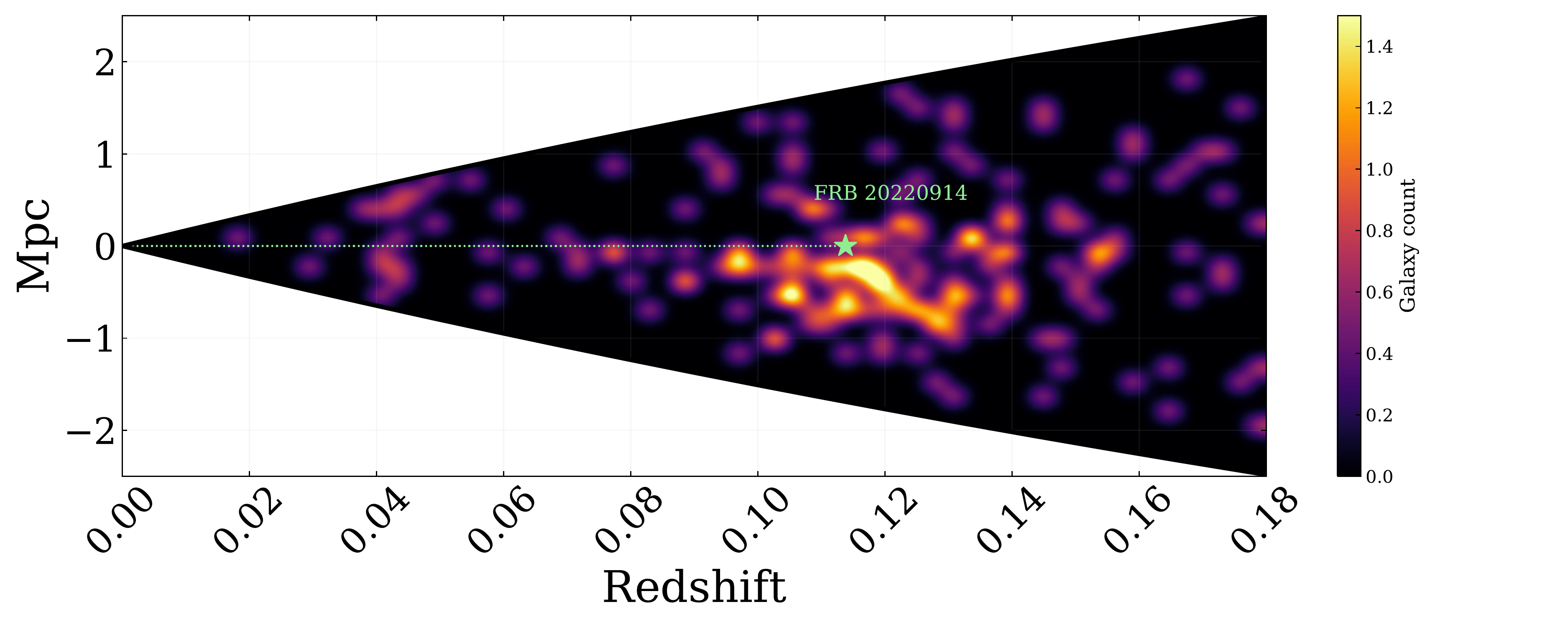}
    \\
    \includegraphics[width=0.5
    \textwidth]{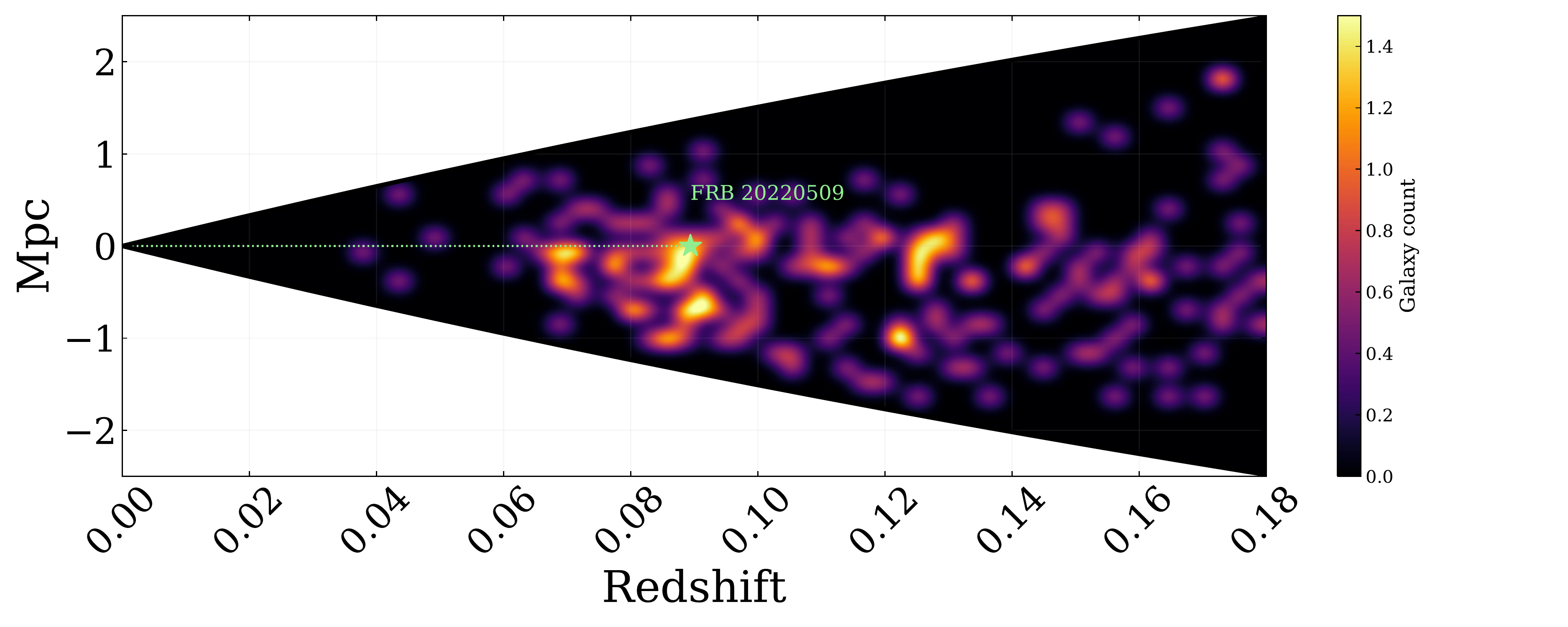}
    \caption{Slice plots of the smoothed galaxy number density within 5 arcminutes of each FRB line-of-sight. The 
    galaxy data are from the DESI Legacy Imaging Survey DR9 and are mostly photometric redshifts.}
    \label{fig:desi-overdensity}
\end{figure}
The burst was fit with a scattering tail plus 
a Gaussian component. We find evidence of scattering 
with a timescale of 80$\pm$20\,$\mu$s at 1.4\,GHz and a Gaussian 
component that is $30\pm10$\,$\mu$s.
Offline polarimetric analysis found that FRB\,20220509G was 
nearly 100$\%$ linearly polarized with an RM of $-111.5\pm1.5$\,rad\,m$^{-3}$ in the observer frame. 
This is significantly larger than the Galactic RM foreground value of
$-9\pm16$\,rad\,m$^{-3}$.
The burst's position angle (PA) is flat across the pulse.
The polarization properties of FRB\,20220509G are 
shown in Figure~\ref{fig:jackiepol}.

\subsubsection{Abell\,2311 galaxy cluster}
The host galaxy of FRB\,20220509G is a member of 
the galaxy cluster Abell\,2311 \citep{Abell}. According to the DESI DR9 cluster/group catalog,
the cluster's richness is 
190 with $M_{180}=2.5\times10^{14}\,M_\odot$, determined by the velocity dispersion of member galaxies. The radius at which 
the average enclosed density is 500 times the critical density is 
$R_{500}=800$\,kpc. 
The X-ray luminosity within $R_{500}$ is 
$L_X=4.7\times10^{43}$\,erg\,s$^{-1}$ with an inferred mass of $M_{500}=1.6\times10^{14}\,M_\odot$. 
The cluster 
does not have a published SZ detection and is not in the Planck SZ2 cluster catalog. However, we have analyzed the Planck MILCA $y$-map for this cluster in the same way as for Abell 2311, finding $Y_{5R500}=0.81\pm0.36 \times 10^{-3}$ arcmin${^2}$.

\begin{figure*}
    \centering
    \includegraphics[width=0.85\textwidth]{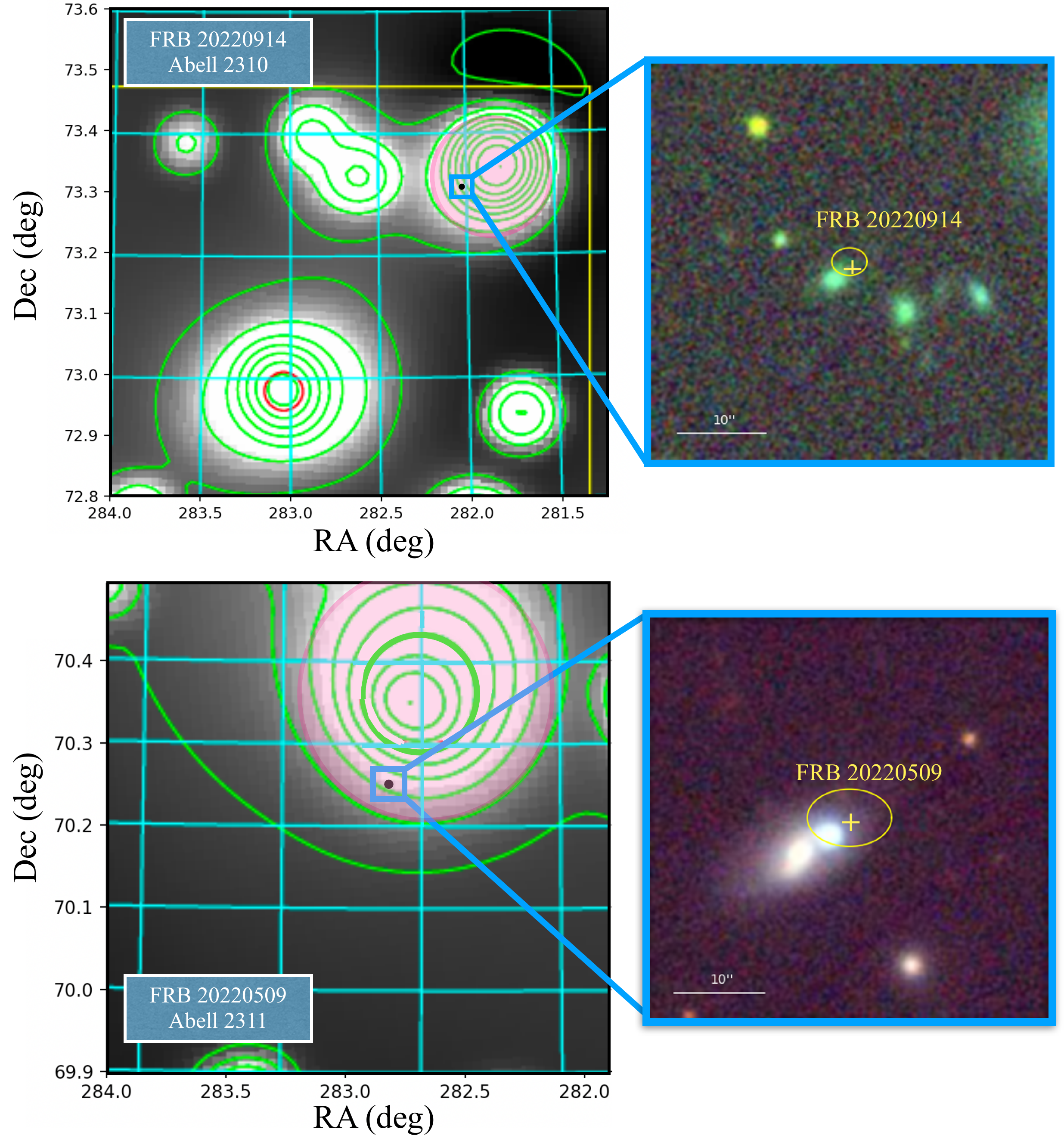}
    \caption{The two FRB sources reported in this work with their respective host galaxies and galaxies clusters. The left column images show filtered ROSAT X-ray intensity \citep{weiwei-rosat} for 
    galaxy cluster Abell\,2310 (top) and Abell\,2311 (bottom). The magenta circle
    in the top left panel
    has a radius of 1\,Mpc and denotes the location of the SZ cluster from the PSZ2 catalog. The bottom panel magenta cirlce corresponds to our SZ measurement for Abell\,2311. The right columns show
    insets on the position of FRB\,20220914A (top) and FRB\,20220509G (bottom) plotted 
    over a PanSTARRS1 riz-band image \citep[PS1]{chambersPS1}.}
    \label{fig:cluster-elektra}
\end{figure*}

The filtered ROSAT X-ray intensity is shown in 
Figure~\ref{fig:cluster-elektra}. The projected physical offset 
between FRB\,20220509G and the X-ray centroid is 
$870\pm 50$\,kpc, placing the FRB at a minimum 
radius of just beyond $R_{500}$.

\subsubsection{The host galaxy of FRB\,20220509G}

The likely host galaxy of FRB 20220509 is an early-type elliptical galaxy with insignificant ongoing star formation, and hence stands out as a quiescent galaxy in the known population of FRB hosts~(Sharma et al., in prep.). The measured spectroscopic redshift of the host is $0.0894\pm0.0001$ and a detailed spectral energy distribution analysis reveals a stellar mass of $\log \mathrm{M}_* (\mathrm{M}_\odot) = 11.13^{+0.02}_{-0.02}$ with star formation rate averaged over the last 100~Myr of $0.08^{+0.06}_{-0.04}$. An old stellar population implies a long delay between the time of occurrence of this FRB and the formation of its progenitor, thus opening another window of progenitors with long delay times.


\section{Data analysis}
We now seek to synthesize the 
FRB radio properties with observations 
of their host galaxies and host clusters.
This will allow us to infer the impact of the 
ICM on FRB\,20220509G and FRB\,20220914A. Neither 
host galaxy has a high rate of star formation. 
The early-type host of FRB\,20220509G, 
in particular, is not expected to have a significant 
ISM that would contribute to 
local dispersion \citep{lees92}. In this section we 
estimate the origin of scattering, DM, and RM 
and model the contribution of the ICM 
to each propagation effect. 

\subsection{Scattering \& scintillation}
Both FRB sources show evidence of scintillation that 
is consistent with a nearby scattering screen 
in the Milky Way. FRB\,20220509G and FRB\,20220914A 
have decorrelation bandwidths of $\Delta\nu=3.1 \pm 0.7$\,MHz and $\Delta\nu=2.51 \pm 0.16$\,MHz, respectively. 
Since the two sources are only 
separated by a few degrees, it is unsurprising 
that their decorrelation bandwidth and 
modulation indexes are similar. However, 
the presence of fully-modulated scintillation is physically interesting in the case of 
FRB\,20220509G, which also shows evidence of 
temporal scattering. The presence of 
both scattering and spectral scintillation 
allows one to place constraints on the geometry 
of the two scattering screens. Scattering causes 
angular broadening and if that broadening were 
too large the FRB would be resolved out by 
the scattering screen in our Galaxy \citep{masui15, connor16, simard21}. 

For a 
given scattering timescale, angular broadening 
is maximal when the screen is halfway between us and the source. Therefore, if the screen were in the CGM of 
an intervening galaxy at $z\sim0.05$, we 
would not expect FRB\,20220509G to scintillate. Following Eq~48 in \citet{simard21}, we place an upper-limit 
on the distance between the FRB source and the first scattering screen to be $\leq140$\,kpc. This leads to three scenarios for the origin of the temporal 
scattering in FRB\,20220509G, each of which is plausible but somewhat surprising 
given the early-type host galaxy and the hot, smooth 
ICM in which it is embedded.
The scattering could be in the ISM of the host galaxy at 
$\sim$\,kpc scales from the FRB emitting source. This would require an unusual 
sightline given the lack of H$\alpha$ emission and limited turbulence 
in the ISM of elliptical galaxies \citep{seta21, ocker2022b}.
Alternatively, it could arise in the immediate vicinity of the source, such as in a stellar wind, analogous to the scattering in FRB\,20190520 \citep{niu190520, ocker190520, beniamini, reshma1905250}. A third option is that it could be near the host galaxy in the ICM, perhaps in the outflows or CGM of the host. 


\subsection{DM from the ICM}
\label{sec:dmbudget}
We model the DM along the line of sight to each FRB in order 
to generate a probability distribution for the ICM contribution. The probability density function (PDF, denoted here 
by $\mathcal{P}$) of 
a sum of independent variables is the convolution of 
their individual PDFs. The DM terms are not strictly independent, 
but we find that the convolution relation 
for a sum of variables is a good approximation 
in this case. Using Eq.~\ref{eq:dmbudget}, 
we find,

\begin{equation}
    \mathcal{P}(\rm{DM}_{ICM})\!=\!\mathcal{P}(DM_{obs})\ast
    \mathcal{P}(-\rm{DM}_{MW})
    \ast
    \mathcal{P}(-\rm{DM}_{IGM})\ast
    \mathcal{P}(-\rm{DM}_{h})
\end{equation}
\vspace{2pt}
\\

\noindent where $\mathcal{P}(\rm{DM}_{obs})$ is 
Gaussian with variance determined by the measurement error on the observed DM. Here the DMs are all in the observer frame, so we must multiply by $1+z_{c}$ when estimating the DM from the ICM. 
We assert that the probability density must be zero 
for negative DM. 

For FRB\,20220914A, we place an upper-limit on 
$\rm{DM}_{h}$ to be 50\,pc\,cm$^{-3}$ based on 
the strong upper limit on temporal scattering.
We take $\mathcal{P}(\rm{DM}_{h})$
to be a uniform distribution between 0 and 50\,pc\,cm$^{-3}$ because we do not have a strong motivation for a preferred DM peak in the distribution. Previous detections of unscattered FRBs 
do not have a robust measurement of local DM on which to base an empirical distribution. There is reason to believe 
the local DM contribution is very small for other similar FRB sources. For example, FRB\,20180916B \citep{inesR3} and FRB\,20220319D \citep{ravi-mark2023} have comparable upper-limits on local 
scattering and likely have DM$_{h}\leq20$\,pc\,cm$^{-3}$. 
The local ($<150$\,kpc) 
contribution to the DM of FRB\,20220509G is 
less certain, due to the presence of both scattering and scintillation. We assume a flat prior between 0 and 100\,pc\,cm$^{-3}$. 

\begin{figure}[h]
    \centering
    \includegraphics[width=0.5\textwidth]{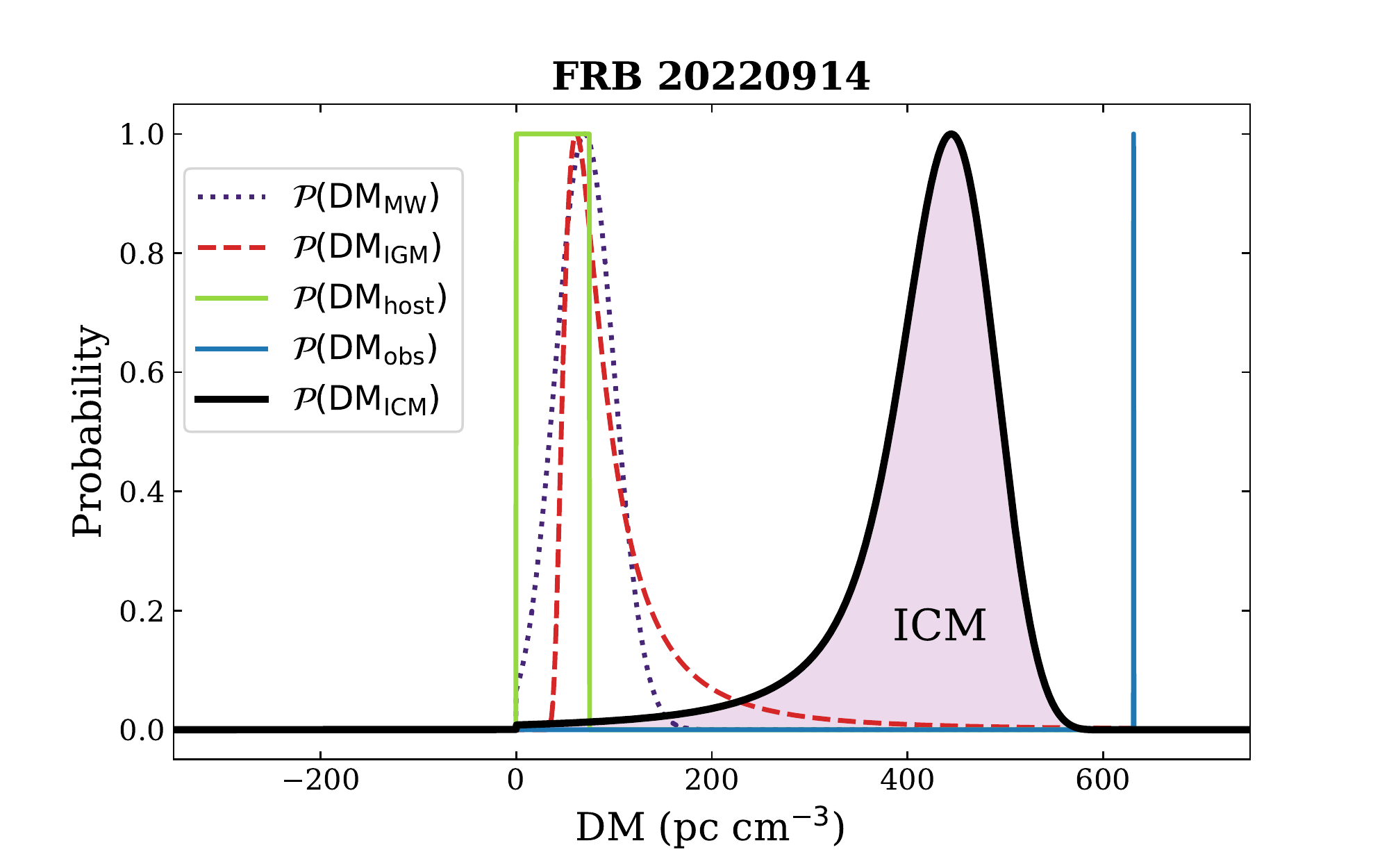}\\
    \includegraphics[width=0.5\textwidth]{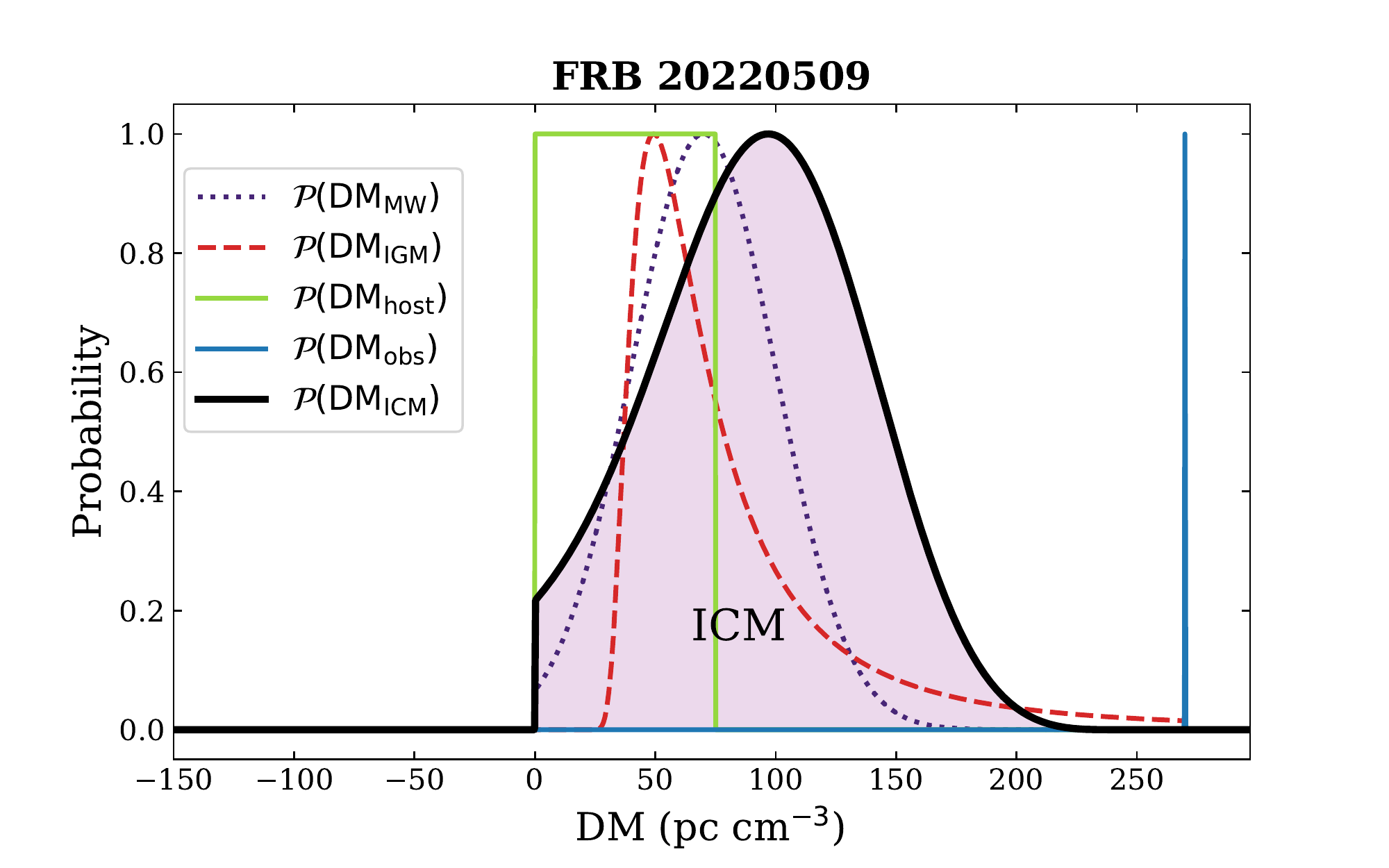}
    \caption{The relative probability 
    distributions for each component of the 
    observed FRB DMs. The distributions 
    of $\rm DM_{\rm ICM}$ are shown in the purple 
    shaded region under the black curves.}
    \label{fig:dmbudget}
\end{figure}

For $\mathcal{P}(\rm{DM}_{IGM})$, we use the 
IllustrisTNG simulation to estimate the distribution 
of DM from the IGM for a source at 
a few hundred Mpc along similar sightlines to our FRBs. We take the 
functional form of DM$_{\rm IGM}$ found by 
\citet{zhangDM2021} for FRBs at $z\approx0.1$, 
who also used IllustrisTNG.
However, we independently estimate the mean 
DM$_{\rm IGM}$ for sightlines 
that intersect massive halos with $M_{500}>10^{13.5}\,M_\odot$. We then exclude DM contribution from the halo itself to estimate the mean of $\rm DM_{\rm IGM}$ for that subset of sightlines. 
By doing this we account for correlations 
in the Universe's matter distribution: Sightlines that intersect clusters are more likely to intersect filaments and less likely to pass through voids, compared to typical positions. We find that for $z\lesssim0.1$, 

\begin{equation}
\langle \rm DM_{IGM} \rangle \approx 22\,\rm pc\,cm^{-3}\,\left (\frac{D_{A}}{100\,Mpc} \right),
\end{equation}

\noindent where $D_{A}$ is angular diameter distance. This gives a mean value of 93\,pc\,cm$^{-3}$ with a 90$\%$ confidence interval 
on DM$_{\rm IGM}=48-260$\,pc\,cm$^{-3}$ for 
FRB\,20220914A. For FRB\,20220509G we find a mean 
value of 76\,pc\,cm$^{-3}$ with a 90$\%$ confidence interval 
on DM$_{\rm IGM}=39-175$\,pc\,cm$^{-3}$. The estimated IGM DM distributions are shown 
in Figure~\ref{fig:dmbudget} as dashed 
red curves.

In these FRB directions, the Milky Way contribution is taken to be 
normally distributed with mean 70\,pc\,cm$^{-3}$ and standard deviation 30\,pc\,cm$^{-3}$.
This value includes the Galactic halo. These values 
are based on NE2001 \citep{ne2001} and YMW \citep{ymw16} and recent evidence that the Milky Way halo contribution to DM is smaller than previously thought \citep{M81, ravi-mark2023, cook23}. The two FRB sources 
are nearby on the sky, allowing us
to use similar estimates for $\mathcal{P}(\rm{DM}_{MW})$.

We have attempted to make conservative estimates 
on the uncertainty of each component of the 
observed DM, which will result in a wider 
inferred DM$_{\rm ICM}$ distribution. We warn that 
the various elements in the DM budget were estimated by different means and are subject to  
modelling uncertainties.
With these caveats, the resulting 
probability distribution of the 
ICM contribution to DM is shown in Figure~\ref{fig:dmbudget}. We find that 
Abell\,2310's ICM adds 265--511\,pc\,cm$^{-3}$
at 90$\%$ confidence to the observed DM of FRB\,20220914A. In the frame of the cluster, 
this range is 295--568\,pc\,cm$^{-3}$. These 
values make it unlikely that the host galaxy 
is near the front of the cluster, as such a 
high $\rm DM_{\rm ICM}$ requires a significant 
path through the intracluster medium. 
In the case of Abell\,2311, we find a 
90$\%$ confidence interval of 16--172\,pc\,cm$^{-3}$
for $\rm DM_{\rm ICM}$ in the cluster frame. 
This range is consistent with an FRB embedded in a $1.6\times10^{14}\,M_\odot$ cluster 
at a projected offset of $\sim$\,870\,kpc. Still, 
it is difficult to say with certainty that 
the FRB's majority DM component is the ICM 
due to the presence of scattering and larger 
projected impact parameter of the host.

\subsection{ICM magnetic field}
FRBs are the only known cosmological sources 
for which both DM and RM can be measured. 
If DM$_{\rm ICM}$ and RM$_{\rm ICM}$ can be determined, 
FRBs allow us to measure the mean line-of-sight 
magnetic field strength of a galaxy cluster, 
as shown in Eq.~\ref{eq:Bperp}. 
The observed RM of FRB\,20220509G is $-111.5\pm1$\,rad\,m$^{-2}$, compared with the expected 
Milky Way foreground RM of $-9\pm16$\,rad\,m$^{-2}$ 
in that direction \citep{rmmapMW}.
If the extragalactic RM is dominated by the ICM then 
the cluster-frame RM is $\approx$\,120\,rad\,m$^{-2}$, 
and we can use $\mathcal{P}(\rm DM_{ICM})$ to infer 
$\langle B_{\parallel, \rm ICM} \rangle$ at 
the impact parameter of FRB\,20220509G. We find that the average magnetic field strength is 0.7--7.5\,$\mu$G. 
Again, this is subject to the assumption that the observed RM is dominated by the ICM. 

The values we obtain for both RM and $\langle B_{\parallel, \rm ICM} \rangle$ are in line with previous observations, as well as magnetohydrodynamical (MHD) simulations of galaxy clusters \citep{clustermagreview, illustriscluster}. \citet{ICMbfield2016} analyzed the RMs from 1383 
of extragalactic polarized sources and found values
between $-200$ and $+$200\,rad\,m$^{-2}$ 
were common at $R_{500}$. The corresponding 
magnetic field strengths they deduced were $\sim$
a few microGauss \citep{ICMbfield2016}. In Section~\ref{sec:discussion} we discuss how cluster FRBs 
from future surveys could better constrain 
magnetic fields in and around the ICM.

\subsection{Mean temperature from DM}

The ICM observables 
that we discuss in this work are 
DM, RM, X-ray brightness, and the SZ-$y$ parameter. Each is a 
weighted integral of $n_e$ along the line of sight. In Faraday rotation,  
the gas density is weighted by magnetic field strength, $B_{//}$. For X-ray luminosity, emission 
is the sum of $n^2_e$ weighted by $\Lambda(T_e)$. In the case 
of the SZ-$y$ parameter, $n_e$ is weighted by $T_e$ itself. 
X-ray derived density and SZ-derived pressure are 
commonly used to estimate temperature \citep{xraySZ}.
Alternatively, the ratio of X-ray luminosity to $y_{SZ}$ has been taken to constrain 
the gas temperature. For example, \citet{PSZvirgo} find the 
ratio of X-ray luminosity to $y_{SZ}$ for the Virgo cluster 
at angular position 
$\theta$ to be $\langle n_e\,\Lambda(T_e) / (k_B\,T_e) \rangle(\theta)$. 
This quantity offers information on the cluster temperature profile 
under certain assumptions about $n_e(r)$ and $\Lambda$.

We can attempt something similar using the ratio of $y_{SZ}$ to 
DM$_{\rm ICM}$, which will be independent of $n_e$. Evaluating Eq.~\ref{eq:sz} and Eq.~\ref{eq:DM} 
at impact parameter $b_\perp$ we get,

\begin{equation}
    y_{SZ}(b_\perp) \approx \frac{k_B\,T_e}{m_e\,c^2}\,\sigma_T\,\langle n_e \rangle\,L_{ICM}
\end{equation}

\begin{equation}
    \textup{DM}_{ICM}(b_\perp) \approx \langle n_e \rangle\,L_{ICM},
\end{equation}

\noindent where $L_{ICM}$ is the line-of-sight lengthscale of the cluster at that 
$b_\perp$. By taking their ratio and re-arranging we can make a 
crude approximation for the mean plasma temperature 
at that physical offset,

\begin{align}
    \langle T_e(b_\perp) \rangle &\approx \frac{m_e\,c^2}{\sigma_T\,k_B}\,\frac{y_{sz}}{\textup{DM}_{ICM}}
    \\
    &=1.15\times10^7\,\mathrm{K}\,
    \left (\frac{y_{SZ}}{10^{-6}} \right )
    \left (\frac{\rm DM}{250\,\frac{\rm pc}{\rm cm^{3}}} \right )^{-1}
\label{eq:dmyratio}
\end{align}

\noindent We estimate the projected $y_{sz}$ at each cluster's $b_\perp$ 
from our fits to the Planck data described in Section~\ref{sec:observations}. This gives 
$y_{SZ} = (3.31\pm1.36)\times10^{-6}$ for Abell\,2310 
at 520\,kpc and $y_{SZ} = (6.80\pm3.03)\times10^{-7}$ for Abell\,2311 at 870\,kpc. 
Plugging these values into 
Eq.~\ref{eq:dmyratio}, we find $\langle k_B\,T_e \rangle \approx 0.8-3.9$\,keV for Abell\,2310,
in agreement with the expected electron temperature 
for a cluster of similar mass at that impact parameter \citep{Vikhlinin2005,Pratt2007}. 
This is the first time halo 
gas temperature has been measured using 
fast radio bursts. For Abell\,2311 at 870\,kpc we find 
$\langle k_B\,T_e \rangle \approx 0.6-16$\,keV 
along that line of sight. The large range in temperature 
is primarily due to the uncertainty in ICM DM 
for FRB\,20220509G. 

While it is hard to match the statistical precision of an X-ray spectroscopic
temperature measurement, 
there are benefits to using FRB DMs to derive gas temperature:
With FRBs we do not need to 
account for ``clumping'' effects due to the 
$n^2_e$ dependence \citep{Eckert2015}, the temperature will be approximately
mass-weighted rather than the more complicated weighting intrinsic to spectroscopic
measurements \citep{Mazzotta2004}, and current X-ray facilities are generally
not sensitive beyond $\sim$\,$R_{500}$. 
The obvious drawbacks of using FRBs are that we have only a single sightline and the DM contribution of the IGM, Milky Way, and the host galaxy must be modelled. The former will be alleviated by FRB surveys with high areal density, providing multiple sightlines through individual halos. The latter may be aided by 
cross-correlation (e.g. \citep{Madhavacheril, chimexcorr}), which are less sensitive to host DM contamination. In 
Section~\ref{sec:discussion} we describe how future surveys will 
allow for statistical FRB/cluster studies.

\section{IllustrisTNG Cluster Simulations} 
\begin{figure*}
    \centering
    \includegraphics[width=1\textwidth]{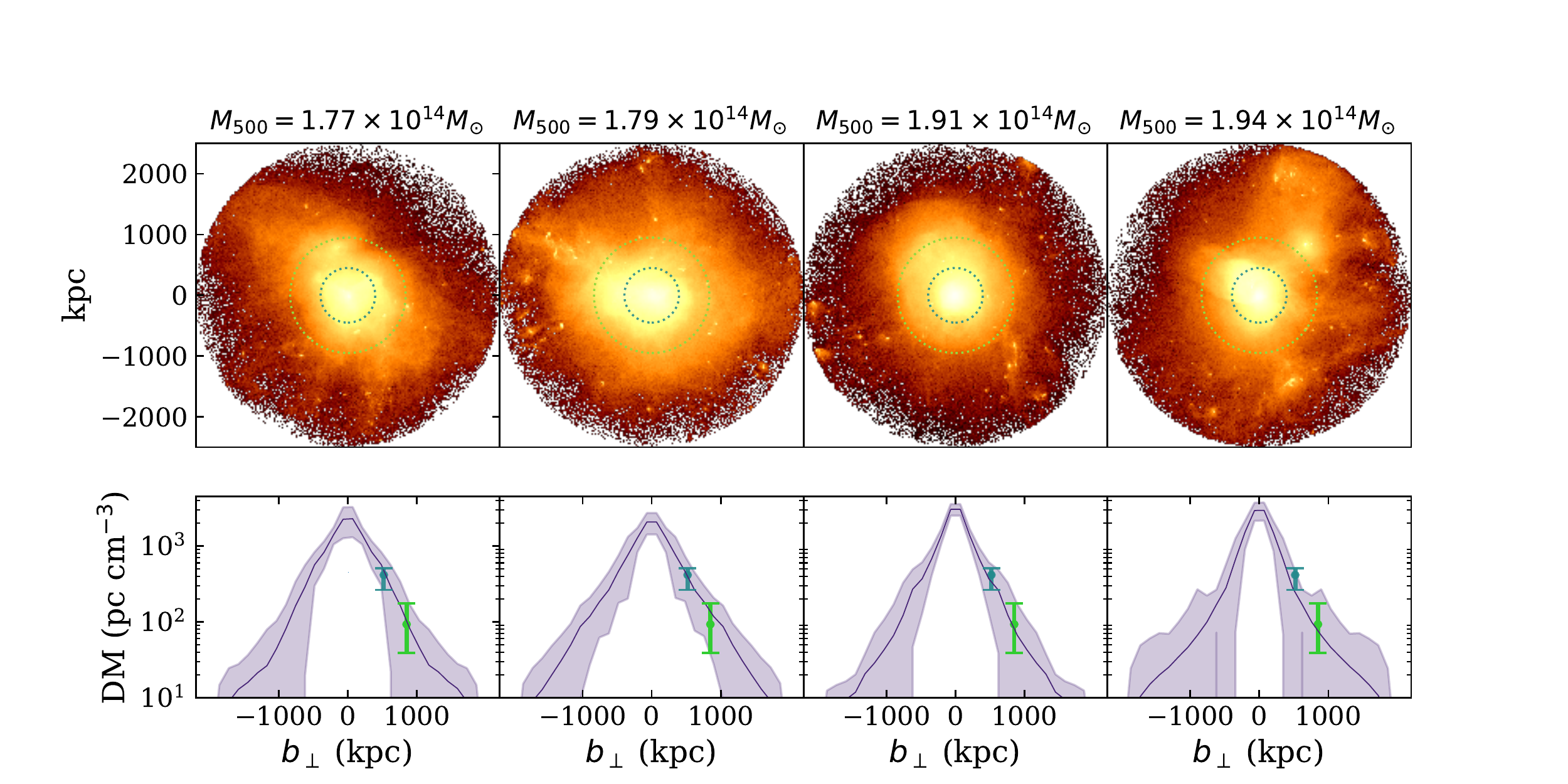}
    \caption{Four galaxy clusters from the TNG300-1 simulation whose 
    masses are similar to those of Abell\,2310 and Abell\,2311. The top 
    row show gas surface density within 2\,Mpc of the cluster center. The green and 
    blue circles represent the impact parameter of FRB\,20220509G and 
    FRB\,20220914A, respectively. The bottom row shows DM$_{\rm ICM}$ as a function 
    of impact parameter, $b_\perp$, with error bars for the two observed cluster FRBs. The solid purple line is 
    the median DM over fifty azimuthal angles and the shaded region 
    shows the $90\%$ limits at that $b_\perp$. The curves assume the 
    FRB is embedded halfway in the ICM, i.e. $l_{LOS}=0$.}
    \label{fig:illustris}
\end{figure*}

We use the IllustrisTNG simulation 
TNG300-1 \citep{illustrisTNG} to estimate typical DM 
values of the ICM as a function of 
impact parameter. We  
seek also to estimate the variance 
between galaxy clusters of similar 
mass, and variance 
within a cluster for different position angle 
at fixed impact parameter. 

We first extract cylinders 
of length 300\,Mpc and radius 5\,Mpc
from the simulation box that contain 
massive halos. These allow us to 
estimate not only the ICM contribution to DM, 
but also the IGM along that sightline. We then 
calculate gas properties for 
each cell in that cylinder and estimate free electron density as,

\begin{equation}
    n_e = f_e\,\rho_g\,X_H\,/m_p,
\end{equation}

\noindent where $f_e$ is the 
free electron abundance, $\rho_g$ is 
gas density, $X_H$ is hydrogen 
abundance, and $m_p$ is the proton mass. We take
$X_H$ to be 0.75. 
The $i^{th}$ cell's contribution to the total DM  is,

\begin{equation}
    \mathrm{DM}_i = n^i_e\,\Delta r_i\,(1+z_i)^{-1},
\end{equation}

\noindent where $\Delta r_i$ is the distance 
through the cell at an impact parameter $b_i$. 
The cell's redshift is 
$z_i$. 
For cellsize, $\Delta x$, 
$\Delta r_i=2\sqrt{\Delta x^2 - b^2_i}$. The total 
DM is then $\Sigma DM_i$ for all cells that 
are intersected by the FRB.

We compute DM profiles for all halos in 
TNG300-1 that have $M_{500}>10^{13.5}\,M_\odot$. In Figure~\ref{fig:illustris}
we show four galaxy clusters with masses comparable to
the masses of host clusters of FRB\,20220509G and FRB\,20220914A. The blue and green 
data points in the bottom row correspond to the 
estimated ICM contribution to the two FRB DMs, 
which agree well with the expected range at their 
respective impact parameters.
There is significant variance within a cluster 
even at a fixed impact parameter, particularly beyond 
the virial radius (corresponding to $\approx1500$\,kpc given 
$R_{500}\approx800$\,kpc for these clusters). There is also some scatter between clusters, 
but the DM curves are roughly consistent. The mean values 
in DM vs. $b_\perp$ (solid purple lines in Figure~\ref{fig:illustris}) are also in agreement with 
analytic models for the cluster gas density \citep{Vikhlinin2005, prochaska2019b}.

Compared to simulations of the CGM or gas in 
galaxy groups, cluster simulations are in relatively 
good agreement with one another as well as with observations \citep{Oppenheimer21}. As the sophistication of cosmological 
MHD simulations progresses, they will be an invaluable tool 
for understanding the properties of FRBs that are impacted 
by cluster gas. This is especially true for magnetic field 
inference from FRB RMs and for gas temperatures outside of $R_{500}$, as outlined in the following section.

\section{Discussion}
\label{sec:discussion}

\subsection{Beyond the virial radius}
Outside of the virial radii of galaxy clusters, plasma is difficult 
to observe via X-ray. FRB DMs, however, are more sensitive to this gas due to the $n_e$  (rather than $n_e^2$) dependence, and can
constrain to ICM and WHIM on the outskirts of clusters. \citet{whimfrb2017} have
shown that by combining FRB DMs with the pressure
profiles from SZ observations, the temperature profile of
the WHIM can be obtained beyond $1.5$\,$R_{200}$.

In future FRB surveys, large numbers of 
localized FRBs will produce DM maps with high areal 
density, intersecting individual 
clusters along multiple sightlines. For example, 
the proposed DSA-2000 survey will spend roughly 
$5\%$ of its time on deep-drilling fields that overlap 
with Rubin, XMM-LSS, Extended Chandra Deep Field-South, COSMOS deep fields \citep{dsa-2000-whitepaper}. This will lead to 
roughly 500 FRBs in just $\sim$\,10\,deg$^2$. Nearly 
every massive galaxy cluster out to $z=0.3$ 
will be intersected by at least one FRB sightline.
We will have a wealth of multiwavelength data in 
this field, 
including SZ data and deep X-ray observations. Such a 
dataset will allow us to constrain the magnetic field 
in the ICM and the density profile in the WHIM, as well as gas temperatures outside of $R_{500}$. However, from 
Figure~\ref{fig:illustris} we see substantial azimuthal and inter-cluster variance in DM at large radii. Analyses of the gas beyond $R_{500}$ ought to be done in tandem with cosmological MHD simulations.

\subsection{Prevalence of cluster FRBs}
Roughly 5$\%$ of sightlines from FRBs at $z\approx0.5$ 
are expected to intersect galaxy clusters in the foreground \citep{prochaska2019b}. 
However, if an FRB sightline has a projected offset from 
a cluster 
that is less than 
$R_{200}$, it is more likely that the FRB came from the cluster itself 
than from behind it, assuming the rate of 
FRBs is related to stellar mass and not just star formation.
This is because the overdensity of galaxies within 
massive halos is enough to counter the larger 
volume behind the galaxy cluster. Using 
$R_{200}\approx1$\,Mpc, the comoving volume 
in region of a cone behind a redshift 0.25 cluster 
out to $z=0.5$ is $\sim$\,$7\times10^{3}$\,Mpc$^3$. 
Assuming a galaxy number density of roughly 
$10^{-2}$\,Mpc$^{-3}$ \citep{gwgc}, the volume behind the 
cluster should have only $\mathcal{O}(70)$ galaxies, compared 
with $10^{2-3}$ galaxies in the cluster. We note again that 
this rough estimate assumes cluster galaxies are as likely 
as non-cluster galaxies to produce FRBs. We have also 
ignored luminosity function considerations, which 
make FRBs behind the cluster to be more difficult to 
detect because they are less bright. It is therefore unsurprising that the first 
two localized FRBs that are known 
to be impacted by the ICM have come from member galaxies of clusters and not from behind clusters.

We have cross-matched twelve sources localized by the DSA-110
with cluster catalogs and found two FRBs that reside in galaxy clusters. 
This suggests the fraction of FRBs 
from galaxy clusters is 
$f_{cFRB}=0.17^{+0.36}_{-0.12}$, using a 90\% Poissonian confidence interval. We do not include other localized FRBs 
because we are not aware of concerted efforts 
to cross-match those FRB positions with clusters. 

The large value of $f_{cFRB}$ has interesting implications 
for FRBs detected at other surveys, and on our interpretation 
of observables such as the FRB DM distribution. 
CHIME/FRB detects 
$\mathcal{O}(10^3)$ FRBs per year, without 
sufficient localization precision to identify a host galaxy. They may therefore already have detected hundreds of sources that reside in galaxy clusters. The host ICM
could then dominate the high-DM tail of the total observed DM distribution \citep{James2022}, which would otherwise be interpreted as the most distant sources. 
Even with arcminute localizations, a careful statistical cross-match of CHIME/FRB sources with galaxy cluster catalogs could reveal this signal.

\subsection{FRB progenitor implications}
FRB\,20220509G is the first source to belong to an early-type galaxy. This fits an emerging picture that 
FRBs can be produced in a variety of environments, 
including dense star forming regions, globular clusters, and pristine environments with little HII \citep{frb121102, marcoteR3, inesR3, franzM81, frb190520}. The full implications of FRB\,20220509G's 
quiescent host will be fleshed out by ~Sharma et al. (in prep.). If more FRBs are localized to cluster galaxies with low rates of star formation, then progenitor models must explain evolutionary channels that can produce FRBs in the absence of 
recent core-collapse supernovae. This 
would be promising for FRB applications research, as early-type galaxies are not 
expected to have significant magnetoionic plasma in their ISM or HII regions near the source, alleviating the problem of disentangling host DM, RM, and scattering from 
cosmological contributions. 

\section{Conclusion}
We have discovered two fast radio bursts sources that reside in massive galaxy clusters. The host galaxy of FRB\,20220914A 
is a member of cluster Abell\,2310, whose ICM dominates the DM budget of the FRB. We have combined the DM of FRB\,20220914A with SZ observations of Abell\,2310 to make the first estimate of the mean line-of-sight temperature of gas using an FRB. FRB\,20220509G belongs to an early-type galaxy~(Sharma et al., in prep.) at a projected offset of 870\,kpc from the 
center of cluster Abell\,2311. That cluster's ICM is found to contribute 16--172\,pc\,cm$^{-3}$ to the FRB DM, which is consistent with analytic models and simulations of the ICM for a cluster with mass $M_{500}\approx1.5-2\times10^{14} M_\odot$. Polarization analysis of the burst found significant 
Faraday rotation. Assuming this RM originates in the ICM, we constrain the mean line-of-sight magnetic field strength to be 0.75-7\,$\mu$G in the intracluster gas. Roughly 17$\%$ of our first sample of localized FRBs were found to reside in galaxy clusters. While we expect this fraction to come down with time, galaxy clusters will likely play a significant role 
in upcoming FRB surveys, particularly in attempts at mapping out the Universe's baryons. 

\begin{acknowledgments}

The authors thank staff members of the Owens Valley Radio Observatory and the Caltech radio group, including Kristen Bernasconi, Stephanie Cha-Ramos, Sarah Harnach, Tom Klinefelter, Lori McGraw, Corey Posner, Andres Rizo, Michael Virgin, Scott White, and Thomas Zentmyer. Their tireless efforts were instrumental to the success of the DSA-110. The DSA-110 is supported by the National Science Foundation Mid-Scale Innovations Program in Astronomical Sciences (MSIP) under grant AST-1836018. We acknowledge use of the VLA calibrator manual and the radio fundamental catalog. Some of the data presented herein were obtained at the W. M. Keck Observatory, which is operated as a scientific partnership among the California Institute of Technology, the University of California and the National Aeronautics and Space Administration. The Observatory was made possible by the generous financial support of the W. M. Keck Foundation. We thank Ryuichi Takahashi and Weiwei Xu for helpful discussions and materials related to their overlapping 
work. We also thank Nicholas Battaglia for useful discussion. 

\end{acknowledgments}

\facility{Keck:I (LRIS), Keck:II (ESI)} 
\software{astropy, CASA, frb, heimdall, lpipe, pPXF, Prospector, wsclean}

\bibliography{main}{}

\begin{thebibliography}{}
\expandafter\ifx\csname natexlab\endcsname\relax\def\natexlab#1{#1}\fi
\providecommand{\url}[1]{\href{#1}{#1}}
\providecommand{\dodoi}[1]{doi:~\href{http://doi.org/#1}{\nolinkurl{#1}}}
\providecommand{\doeprint}[1]{\href{http://ascl.net/#1}{\nolinkurl{http://ascl.net/#1}}}
\providecommand{\doarXiv}[1]{\href{https://arxiv.org/abs/#1}{\nolinkurl{https://arxiv.org/abs/#1}}}

\bibitem[{chi(2021)}]{chimexcorr}
 2021, 922, 42, \dodoi{10.3847/1538-4357/ac1dab}

\bibitem[{{Abell}(1958)}]{Abell}
{Abell}, G.~O. 1958, \apjs, 3, 211, \dodoi{10.1086/190036}

\bibitem[{{Agarwal} {et~al.}(2019){Agarwal}, {Lorimer}, {Fialkov}, {Bannister},
  {Shannon}, {Farah}, {Bhandari}, {Macquart}, {Flynn}, {Pignata}, {Tejos},
  {Gregg}, {Os{\l}owski}, {Rajwade}, {Mickaliger}, {Stappers}, {Li}, {Zhu},
  {Qian}, {Yue}, {Wang}, \& {Loeb}}]{Devansh2019}
{Agarwal}, D., {Lorimer}, D.~R., {Fialkov}, A., {et~al.} 2019, \mnras, 490, 1,
  \dodoi{10.1093/mnras/stz2574}

\bibitem[{{Anna-Thomas} {et~al.}(2022){Anna-Thomas}, {Connor}, {Burke-Spolaor},
  {Beniamini}, {Aggarwal}, {Law}, {Lynch}, {Li}, {Feng}, {Ocker}, {Cruces},
  {Chatterjee}, {Yu}, {Niu}, \& {Xue}}]{reshma1905250}
{Anna-Thomas}, R., {Connor}, L., {Burke-Spolaor}, S., {et~al.} 2022, arXiv
  e-prints, arXiv:2202.11112.
\newblock \doarXiv{2202.11112}

\bibitem[{{Arnaud} {et~al.}(2010){Arnaud}, {Pratt}, {Piffaretti},
  {B{\"o}hringer}, {Croston}, \& {Pointecouteau}}]{Arnaud2010}
{Arnaud}, M., {Pratt}, G.~W., {Piffaretti}, R., {et~al.} 2010, \aap, 517, A92,
  \dodoi{10.1051/0004-6361/20091341610.48550/arXiv.0910.1234}

\bibitem[{{Beniamini} {et~al.}(2022){Beniamini}, {Kumar}, \&
  {Narayan}}]{beniamini}
{Beniamini}, P., {Kumar}, P., \& {Narayan}, R. 2022, \mnras, 510, 4654,
  \dodoi{10.1093/mnras/stab3730}

\bibitem[{{Bhandari} {et~al.}(2020){Bhandari}, {Sadler}, {Prochaska}, {Simha},
  {Ryder}, {Marnoch}, {Bannister}, {Macquart}, {Flynn}, {Shannon}, {Tejos},
  {Corro-Guerra}, {Day}, {Deller}, {Ekers}, {Lopez}, {Mahony}, {Nu{\~n}ez}, \&
  {Phillips}}]{bhandari2020}
{Bhandari}, S., {Sadler}, E.~M., {Prochaska}, J.~X., {et~al.} 2020, \apjl, 895,
  L37, \dodoi{10.3847/2041-8213/ab672e}

\bibitem[{{Bhardwaj} {et~al.}(2021){Bhardwaj}, {Gaensler}, {Kaspi},
  {Landecker}, {Mckinven}, {Michilli}, {Pleunis}, {Tendulkar}, {Andersen},
  {Boyle}, {Cassanelli}, {Chawla}, {Cook}, {Dobbs}, {Fonseca}, {Kaczmarek},
  {Leung}, {Masui}, {Mnchmeyer}, {Ng}, {Rafiei-Ravandi}, {Scholz}, {Shin},
  {Smith}, {Stairs}, \& {Zwaniga}}]{M81}
{Bhardwaj}, M., {Gaensler}, B.~M., {Kaspi}, V.~M., {et~al.} 2021, \apjl, 910,
  L18, \dodoi{10.3847/2041-8213/abeaa6}

\bibitem[{{B{\"o}hringer} {et~al.}(2016){B{\"o}hringer}, {Chon}, \&
  {Kronberg}}]{ICMbfield2016}
{B{\"o}hringer}, H., {Chon}, G., \& {Kronberg}, P.~P. 2016, \aap, 596, A22,
  \dodoi{10.1051/0004-6361/201628873}

\bibitem[{{B{\"o}hringer} \& {Werner}(2010)}]{x-ray-temp}
{B{\"o}hringer}, H., \& {Werner}, N. 2010, \aapr, 18, 127,
  \dodoi{10.1007/s00159-009-0023-3}

\bibitem[{{Budzynski} {et~al.}(2012){Budzynski}, {Koposov}, {McCarthy},
  {McGee}, \& {Belokurov}}]{galdist}
{Budzynski}, J.~M., {Koposov}, S.~E., {McCarthy}, I.~G., {McGee}, S.~L., \&
  {Belokurov}, V. 2012, \mnras, 423, 104,
  \dodoi{10.1111/j.1365-2966.2012.20663.x}

\bibitem[{{Carilli} \& {Taylor}(2002)}]{clustermagreview}
{Carilli}, C.~L., \& {Taylor}, G.~B. 2002, \araa, 40, 319,
  \dodoi{10.1146/annurev.astro.40.060401.093852}

\bibitem[{{Chambers} {et~al.}(2016){Chambers}, {Magnier}, {Metcalfe},
  {Flewelling}, {Huber}, {Waters}, {Denneau}, {Draper}, {Farrow}, {Finkbeiner},
  {Holmberg}, {Koppenhoefer}, {Price}, {Rest}, {Saglia}, {Schlafly}, {Smartt},
  {Sweeney}, {Wainscoat}, {Burgett}, {Chastel}, {Grav}, {Heasley}, {Hodapp},
  {Jedicke}, {Kaiser}, {Kudritzki}, {Luppino}, {Lupton}, {Monet}, {Morgan},
  {Onaka}, {Shiao}, {Stubbs}, {Tonry}, {White}, {Ba{\~n}ados}, {Bell},
  {Bender}, {Bernard}, {Boegner}, {Boffi}, {Botticella}, {Calamida},
  {Casertano}, {Chen}, {Chen}, {Cole}, {Deacon}, {Frenk}, {Fitzsimmons},
  {Gezari}, {Gibbs}, {Goessl}, {Goggia}, {Gourgue}, {Goldman}, {Grant},
  {Grebel}, {Hambly}, {Hasinger}, {Heavens}, {Heckman}, {Henderson}, {Henning},
  {Holman}, {Hopp}, {Ip}, {Isani}, {Jackson}, {Keyes}, {Koekemoer}, {Kotak},
  {Le}, {Liska}, {Long}, {Lucey}, {Liu}, {Martin}, {Masci}, {McLean}, {Mindel},
  {Misra}, {Morganson}, {Murphy}, {Obaika}, {Narayan}, {Nieto-Santisteban},
  {Norberg}, {Peacock}, {Pier}, {Postman}, {Primak}, {Rae}, {Rai}, {Riess},
  {Riffeser}, {Rix}, {R{\"o}ser}, {Russel}, {Rutz}, {Schilbach}, {Schultz},
  {Scolnic}, {Strolger}, {Szalay}, {Seitz}, {Small}, {Smith}, {Soderblom},
  {Taylor}, {Thomson}, {Taylor}, {Thakar}, {Thiel}, {Thilker}, {Unger},
  {Urata}, {Valenti}, {Wagner}, {Walder}, {Walter}, {Watters}, {Werner},
  {Wood-Vasey}, \& {Wyse}}]{chambersPS1}
{Chambers}, K.~C., {Magnier}, E.~A., {Metcalfe}, N., {et~al.} 2016, arXiv
  e-prints, arXiv:1612.05560, \dodoi{10.48550/arXiv.1612.05560}

\bibitem[{{Chatterjee} {et~al.}(2017){Chatterjee}, {Law}, {Wharton},
  {Burke-Spolaor}, {Hessels}, {Bower}, {Cordes}, {Tendulkar}, {Bassa},
  {Demorest}, {Butler}, {Seymour}, {Scholz}, {Abruzzo}, {Bogdanov}, {Kaspi},
  {Keimpema}, {Lazio}, {Marcote}, {McLaughlin}, {Paragi}, {Ransom}, {Rupen},
  {Spitler}, \& {van Langevelde}}]{chatterjee17}
{Chatterjee}, S., {Law}, C.~J., {Wharton}, R.~S., {et~al.} 2017, \nat, 541, 58,
  \dodoi{10.1038/nature20797}

\bibitem[{{CHIME/FRB Collaboration} {et~al.}(2019){CHIME/FRB Collaboration},
  {Andersen}, {Bandura}, {Bhardwaj}, {Boubel}, {Boyce}, {Boyle}, {Brar},
  {Cassanelli}, {Chawla}, {Cubranic}, {Deng}, {Dobbs}, {Fandino}, {Fonseca},
  {Gaensler}, {Gilbert}, {Giri}, {Good}, {Halpern}, {Hill}, {Hinshaw},
  {H{\"o}fer}, {Josephy}, {Kaspi}, {Kothes}, {Landecker}, {Lang}, {Li}, {Lin},
  {Masui}, {Mena-Parra}, {Merryfield}, {Mckinven}, {Michilli}, {Milutinovic},
  {Naidu}, {Newburgh}, {Ng}, {Patel}, {Pen}, {Pinsonneault-Marotte}, {Pleunis},
  {Rafiei-Ravandi}, {Rahman}, {Ransom}, {Renard}, {Scholz}, {Siegel}, {Singh},
  {Smith}, {Stairs}, {Tendulkar}, {Tretyakov}, {Vanderlinde}, {Yadav}, \&
  {Zwaniga}}]{anderson-chime}
{CHIME/FRB Collaboration}, {Andersen}, B.~C., {Bandura}, K., {et~al.} 2019,
  \apjl, 885, L24, \dodoi{10.3847/2041-8213/ab4a80}

\bibitem[{{CHIME/FRB Collaboration} {et~al.}(2021){CHIME/FRB Collaboration},
  {Amiri}, {Andersen}, {Bandura}, {Berger}, {Bhardwaj}, {Boyce}, {Boyle},
  {Brar}, {Breitman}, {Cassanelli}, {Chawla}, {Chen}, {Cliche}, {Cook},
  {Cubranic}, {Curtin}, {Deng}, {Dobbs}, {Dong}, {Eadie}, {Fandino}, {Fonseca},
  {Gaensler}, {Giri}, {Good}, {Halpern}, {Hill}, {Hinshaw}, {Josephy},
  {Kaczmarek}, {Kader}, {Kania}, {Kaspi}, {Landecker}, {Lang}, {Leung}, {Li},
  {Lin}, {Masui}, {McKinven}, {Mena-Parra}, {Merryfield}, {Meyers}, {Michilli},
  {Milutinovic}, {Mirhosseini}, {M{\"u}nchmeyer}, {Naidu}, {Newburgh}, {Ng},
  {Patel}, {Pen}, {Petroff}, {Pinsonneault-Marotte}, {Pleunis},
  {Rafiei-Ravandi}, {Rahman}, {Ransom}, {Renard}, {Sanghavi}, {Scholz}, {Shaw},
  {Shin}, {Siegel}, {Sikora}, {Singh}, {Smith}, {Stairs}, {Tan}, {Tendulkar},
  {Vanderlinde}, {Wang}, {Wulf}, \& {Zwaniga}}]{chime-catalog1}
{CHIME/FRB Collaboration}, {Amiri}, M., {Andersen}, B.~C., {et~al.} 2021,
  \apjs, 257, 59, \dodoi{10.3847/1538-4365/ac33ab}

\bibitem[{{Connor} \& {Ravi}(2022)}]{connorravi}
{Connor}, L., \& {Ravi}, V. 2022, Nature Astronomy, 6, 1035,
  \dodoi{10.1038/s41550-022-01719-7}

\bibitem[{{Connor} {et~al.}(2016){Connor}, {Sievers}, \& {Pen}}]{connor16}
{Connor}, L., {Sievers}, J., \& {Pen}, U.-L. 2016, \mnras, 458, L19,
  \dodoi{10.1093/mnrasl/slv124}

\bibitem[{{Cook} {et~al.}(2023){Cook}, {Bhardwaj}, {Gaensler}, {Scholz},
  {Eadie}, {Hill}, {Kaspi}, {Masui}, {Curtin}, {Dong}, {Fonseca},
  {Herrera-Martin}, {Kaczmarek}, {Lanman}, {Lazda}, {Leung}, {Meyers},
  {Michilli}, {Pandhi}, {Pearlman}, {Pleunis}, {Ransom}, {Rahman}, {Sand},
  {Shin}, {Smith}, {Stairs}, \& {Stenning}}]{cook23}
{Cook}, A.~M., {Bhardwaj}, M., {Gaensler}, B.~M., {et~al.} 2023, arXiv
  e-prints, arXiv:2301.03502, \dodoi{10.48550/arXiv.2301.03502}

\bibitem[{{Cordes} \& {Chatterjee}(2019)}]{cordes-review}
{Cordes}, J.~M., \& {Chatterjee}, S. 2019, \araa, 57, 417,
  \dodoi{10.1146/annurev-astro-091918-104501}

\bibitem[{{Cordes} \& {Lazio}(2002)}]{ne2001}
{Cordes}, J.~M., \& {Lazio}, T.~J.~W. 2002, arXiv e-prints, astro.
\newblock \doarXiv{astro-ph/0207156}

\bibitem[{{Dey} {et~al.}(2019){Dey}, {Schlegel}, {Lang}, {Blum}, {Burleigh},
  {Fan}, {Findlay}, {Finkbeiner}, {Herrera}, {Juneau}, {Landriau}, {Levi},
  {McGreer}, {Meisner}, {Myers}, {Moustakas}, {Nugent}, {Patej}, {Schlafly},
  {Walker}, {Valdes}, {Weaver}, {Y{\`e}che}, {Zou}, {Zhou}, {Abareshi},
  {Abbott}, {Abolfathi}, {Aguilera}, {Alam}, {Allen}, {Alvarez}, {Annis},
  {Ansarinejad}, {Aubert}, {Beechert}, {Bell}, {BenZvi}, {Beutler}, {Bielby},
  {Bolton}, {Brice{\~n}o}, {Buckley-Geer}, {Butler}, {Calamida}, {Carlberg},
  {Carter}, {Casas}, {Castander}, {Choi}, {Comparat}, {Cukanovaite}, {Delubac},
  {DeVries}, {Dey}, {Dhungana}, {Dickinson}, {Ding}, {Donaldson}, {Duan},
  {Duckworth}, {Eftekharzadeh}, {Eisenstein}, {Etourneau}, {Fagrelius},
  {Farihi}, {Fitzpatrick}, {Font-Ribera}, {Fulmer}, {G{\"a}nsicke},
  {Gaztanaga}, {George}, {Gerdes}, {Gontcho}, {Gorgoni}, {Green}, {Guy},
  {Harmer}, {Hernandez}, {Honscheid}, {Huang}, {James}, {Jannuzi}, {Jiang},
  {Joyce}, {Karcher}, {Karkar}, {Kehoe}, {Kneib}, {Kueter-Young}, {Lan},
  {Lauer}, {Le Guillou}, {Le Van Suu}, {Lee}, {Lesser}, {Perreault Levasseur},
  {Li}, {Mann}, {Marshall}, {Mart{\'\i}nez-V{\'a}zquez}, {Martini}, {du Mas des
  Bourboux}, {McManus}, {Meier}, {M{\'e}nard}, {Metcalfe},
  {Mu{\~n}oz-Guti{\'e}rrez}, {Najita}, {Napier}, {Narayan}, {Newman}, {Nie},
  {Nord}, {Norman}, {Olsen}, {Paat}, {Palanque-Delabrouille}, {Peng},
  {Poppett}, {Poremba}, {Prakash}, {Rabinowitz}, {Raichoor}, {Rezaie},
  {Robertson}, {Roe}, {Ross}, {Ross}, {Rudnick}, {Safonova}, {Saha},
  {S{\'a}nchez}, {Savary}, {Schweiker}, {Scott}, {Seo}, {Shan}, {Silva},
  {Slepian}, {Soto}, {Sprayberry}, {Staten}, {Stillman}, {Stupak}, {Summers},
  {Sien Tie}, {Tirado}, {Vargas-Maga{\~n}a}, {Vivas}, {Wechsler}, {Williams},
  {Yang}, {Yang}, {Yapici}, {Zaritsky}, {Zenteno}, {Zhang}, {Zhang}, {Zhou}, \&
  {Zhou}}]{desi}
{Dey}, A., {Schlegel}, D.~J., {Lang}, D., {et~al.} 2019, \aj, 157, 168,
  \dodoi{10.3847/1538-3881/ab089d}

\bibitem[{{Eckert} {et~al.}(2013){Eckert}, {Molendi}, {Vazza}, {Ettori}, \&
  {Paltani}}]{xraySZ}
{Eckert}, D., {Molendi}, S., {Vazza}, F., {Ettori}, S., \& {Paltani}, S. 2013,
  \aap, 551, A22, \dodoi{10.1051/0004-6361/201220402}

\bibitem[{{Eckert} {et~al.}(2015){Eckert}, {Roncarelli}, {Ettori}, {Molendi},
  {Vazza}, {Gastaldello}, \& {Rossetti}}]{Eckert2015}
{Eckert}, D., {Roncarelli}, M., {Ettori}, S., {et~al.} 2015, \mnras, 447, 2198,
  \dodoi{10.1093/mnras/stu2590}

\bibitem[{{Eckert} {et~al.}(2019){Eckert}, {Ghirardini}, {Ettori}, {Rasia},
  {Biffi}, {Pointecouteau}, {Rossetti}, {Molendi}, {Vazza}, {Gastaldello},
  {Gaspari}, {De Grandi}, {Ghizzardi}, {Bourdin}, {Tchernin}, \&
  {Roncarelli}}]{Eckert2019}
{Eckert}, D., {Ghirardini}, V., {Ettori}, S., {et~al.} 2019, \aap, 621, A40,
  \dodoi{10.1051/0004-6361/201833324}

\bibitem[{{Fujita} {et~al.}(2017){Fujita}, {Akahori}, {Umetsu}, {Sarazin}, \&
  {Wong}}]{whimfrb2017}
{Fujita}, Y., {Akahori}, T., {Umetsu}, K., {Sarazin}, C.~L., \& {Wong}, K.-W.
  2017, \apj, 834, 13, \dodoi{10.3847/1538-4357/834/1/13}

\bibitem[{{Ghirardini} {et~al.}(2019){Ghirardini}, {Eckert}, {Ettori},
  {Pointecouteau}, {Molendi}, {Gaspari}, {Rossetti}, {De Grandi}, {Roncarelli},
  {Bourdin}, {Mazzotta}, {Rasia}, \& {Vazza}}]{2019A&A...621A..41G}
{Ghirardini}, V., {Eckert}, D., {Ettori}, S., {et~al.} 2019, \aap, 621, A41,
  \dodoi{10.1051/0004-6361/201833325}

\bibitem[{{Hallinan} {et~al.}(2019{\natexlab{a}}){Hallinan}, {Dong}, \&
  {Ravi}}]{hallinan-coma}
{Hallinan}, G., {Dong}, D., \& {Ravi}, V. 2019{\natexlab{a}}, The Astronomer's
  Telegram, 13018, 1

\bibitem[{{Hallinan} {et~al.}(2019{\natexlab{b}}){Hallinan}, {Ravi}, {Weinreb},
  {Kocz}, {Huang}, {Woody}, {Lamb}, {D'Addario}, {Catha}, {Law}, {Kulkarni},
  {Phinney}, {Eastwood}, {Bouman}, {McLaughlin}, {Ransom}, {Siemens}, {Cordes},
  {Lynch}, {Kaplan}, {Brazier}, {Bhatnagar}, {Myers}, {Walter}, \&
  {Gaensler}}]{dsa-2000-whitepaper}
{Hallinan}, G., {Ravi}, V., {Weinreb}, S., {et~al.} 2019{\natexlab{b}}, in
  Bulletin of the American Astronomical Society, Vol.~51, 255.
\newblock \doarXiv{1907.07648}

\bibitem[{{Heintz} {et~al.}(2020){Heintz}, {Prochaska}, {Simha}, {Platts},
  {Fong}, {Tejos}, {Ryder}, {Aggerwal}, {Bhandari}, {Day}, {Deller},
  {Kilpatrick}, {Law}, {Macquart}, {Mannings}, {Marnoch}, {Sadler}, \&
  {Shannon}}]{heintz2020}
{Heintz}, K.~E., {Prochaska}, J.~X., {Simha}, S., {et~al.} 2020, \apj, 903,
  152, \dodoi{10.3847/1538-4357/abb6fb}

\bibitem[{{Hitomi Collaboration} {et~al.}(2016){Hitomi Collaboration},
  {Aharonian}, {Akamatsu}, {Akimoto}, {Allen}, {Anabuki}, {Angelini}, {Arnaud},
  {Audard}, {Awaki}, {Axelsson}, {Bamba}, {Bautz}, {Blandford}, {Brenneman},
  {Brown}, {Bulbul}, {Cackett}, {Chernyakova}, {Chiao}, {Coppi}, {Costantini},
  {de Plaa}, {den Herder}, {Done}, {Dotani}, {Ebisawa}, {Eckart}, {Enoto},
  {Ezoe}, {Fabian}, {Ferrigno}, {Foster}, {Fujimoto}, {Fukazawa}, {Furuzawa},
  {Galeazzi}, {Gallo}, {Gandhi}, {Giustini}, {Goldwurm}, {Gu}, {Guainazzi},
  {Haba}, {Hagino}, {Hamaguchi}, {Harrus}, {Hatsukade}, {Hayashi}, {Hayashi},
  {Hayashida}, {Hiraga}, {Hornschemeier}, {Hoshino}, {Hughes}, {Iizuka},
  {Inoue}, {Inoue}, {Ishibashi}, {Ishida}, {Ishikawa}, {Ishisaki}, {Itoh},
  {Iyomoto}, {Kaastra}, {Kallman}, {Kamae}, {Kara}, {Kataoka}, {Katsuda},
  {Katsuta}, {Kawaharada}, {Kawai}, {Kelley}, {Khangulyan}, {Kilbourne},
  {King}, {Kitaguchi}, {Kitamoto}, {Kitayama}, {Kohmura}, {Kokubun}, {Koyama},
  {Koyama}, {Kretschmar}, {Krimm}, {Kubota}, {Kunieda}, {Laurent}, {Lebrun},
  {Lee}, {Leutenegger}, {Limousin}, {Loewenstein}, {Long}, {Lumb}, {Madejski},
  {Maeda}, {Maier}, {Makishima}, {Markevitch}, {Matsumoto}, {Matsushita},
  {McCammon}, {McNamara}, {Mehdipour}, {Miller}, {Miller}, {Mineshige},
  {Mitsuda}, {Mitsuishi}, {Miyazawa}, {Mizuno}, {Mori}, {Mori}, {Moseley},
  {Mukai}, {Murakami}, {Murakami}, {Mushotzky}, {Nagino}, {Nakagawa},
  {Nakajima}, {Nakamori}, {Nakano}, {Nakashima}, {Nakazawa}, {Nobukawa},
  {Noda}, {Nomachi}, {O'Dell}, {Odaka}, {Ohashi}, {Ohno}, {Okajima}, {Ota},
  {Ozaki}, {Paerels}, {Paltani}, {Parmar}, {Petre}, {Pinto}, {Pohl}, {Porter},
  {Pottschmidt}, {Ramsey}, {Reynolds}, {Russell}, {Safi-Harb}, {Saito},
  {Sakai}, {Sameshima}, {Sato}, {Sato}, {Sato}, {Sawada}, {Schartel},
  {Serlemitsos}, {Seta}, {Shidatsu}, {Simionescu}, {Smith}, {Soong}, {Stawarz},
  {Sugawara}, {Sugita}, {Szymkowiak}, {Tajima}, {Takahashi}, {Takahashi},
  {Takeda}, {Takei}, {Tamagawa}, {Tamura}, {Tamura}, {Tanaka}, {Tanaka},
  {Tanaka}, {Tashiro}, {Tawara}, {Terada}, {Terashima}, {Tombesi}, {Tomida},
  {Tsuboi}, {Tsujimoto}, {Tsunemi}, {Tsuru}, {Uchida}, {Uchiyama}, {Uchiyama},
  {Ueda}, {Ueda}, {Ueno}, {Uno}, {Urry}, {Ursino}, {de Vries}, {Watanabe},
  {Werner}, {Wik}, {Wilkins}, {Williams}, {Yamada}, {Yamaguchi}, {Yamaoka},
  {Yamasaki}, {Yamauchi}, {Yamauchi}, {Yaqoob}, {Yatsu}, {Yonetoku}, {Yoshida},
  {Yuasa}, {Zhuravleva}, \& {Zoghbi}}]{x-ray-velocity}
{Hitomi Collaboration}, {Aharonian}, F., {Akamatsu}, H., {et~al.} 2016, \nat,
  535, 117, \dodoi{10.1038/nature18627}

\bibitem[{{Hutschenreuter} {et~al.}(2022){Hutschenreuter}, {Anderson}, {Betti},
  {Bower}, {Brown}, {Br{\"u}ggen}, {Carretti}, {Clarke}, {Clegg}, {Costa},
  {Croft}, {Van Eck}, {Gaensler}, {de Gasperin}, {Haverkorn}, {Heald}, {Hull},
  {Inoue}, {Johnston-Hollitt}, {Kaczmarek}, {Law}, {Ma}, {MacMahon}, {Mao},
  {Riseley}, {Roy}, {Shanahan}, {Shimwell}, {Stil}, {Sobey}, {O'Sullivan},
  {Tasse}, {Vacca}, {Vernstrom}, {Williams}, {Wright}, \&
  {En{\ss}lin}}]{rmmapMW}
{Hutschenreuter}, S., {Anderson}, C.~S., {Betti}, S., {et~al.} 2022, \aap, 657,
  A43, \dodoi{10.1051/0004-6361/202140486}

\bibitem[{{James} {et~al.}(2022){James}, {Prochaska}, {Macquart},
  {North-Hickey}, {Bannister}, \& {Dunning}}]{James2022}
{James}, C.~W., {Prochaska}, J.~X., {Macquart}, J.~P., {et~al.} 2022, \mnras,
  509, 4775, \dodoi{10.1093/mnras/stab3051}

\bibitem[{{Kirsten} {et~al.}(2022){Kirsten}, {Marcote}, {Nimmo}, {Hessels},
  {Bhardwaj}, {Tendulkar}, {Keimpema}, {Yang}, {Snelders}, {Scholz},
  {Pearlman}, {Law}, {Peters}, {Giroletti}, {Paragi}, {Bassa}, {Hewitt},
  {Bach}, {Bezrukovs}, {Burgay}, {Buttaccio}, {Conway}, {Corongiu}, {Feiler},
  {Forss{\'e}n}, {Gawro{\'n}ski}, {Karuppusamy}, {Kharinov}, {Lindqvist},
  {Maccaferri}, {Melnikov}, {Ould-Boukattine}, {Possenti}, {Surcis}, {Wang},
  {Yuan}, {Aggarwal}, {Anna-Thomas}, {Bower}, {Blaauw}, {Burke-Spolaor},
  {Cassanelli}, {Clarke}, {Fonseca}, {Gaensler}, {Gopinath}, {Kaspi}, {Kassim},
  {Lazio}, {Leung}, {Li}, {Lin}, {Masui}, {Mckinven}, {Michilli}, {Mikhailov},
  {Ng}, {Orbidans}, {Pen}, {Petroff}, {Rahman}, {Ransom}, {Shin}, {Smith},
  {Stairs}, \& {Vlemmings}}]{franzM81}
{Kirsten}, F., {Marcote}, B., {Nimmo}, K., {et~al.} 2022, \nat, 602, 585,
  \dodoi{10.1038/s41586-021-04354-w}

\bibitem[{{Lee} {et~al.}(2022){Lee}, {Ata}, {Khrykin}, {Huang}, {Prochaska},
  {Cooke}, {Zhang}, \& {Batten}}]{kglee}
{Lee}, K.-G., {Ata}, M., {Khrykin}, I.~S., {et~al.} 2022, \apj, 928, 9,
  \dodoi{10.3847/1538-4357/ac4f62}

\bibitem[{Lees(1992)}]{lees92}
Lees, J.~F. 1992, Publications of the Astronomical Society of the Pacific, 104,
  154, \dodoi{10.1086/132971}

\bibitem[{{Macquart} {et~al.}(2020){Macquart}, {Prochaska}, {McQuinn},
  {Bannister}, {Bhandari}, {Day}, {Deller}, {Ekers}, {James}, {Marnoch},
  {Os{\l}owski}, {Phillips}, {Ryder}, {Scott}, {Shannon}, \&
  {Tejos}}]{macquart-dm}
{Macquart}, J.~P., {Prochaska}, J.~X., {McQuinn}, M., {et~al.} 2020, \nat, 581,
  391, \dodoi{10.1038/s41586-020-2300-2}

\bibitem[{{Madhavacheril} {et~al.}(2019){Madhavacheril}, {Battaglia}, {Smith},
  \& {Sievers}}]{Madhavacheril}
{Madhavacheril}, M.~S., {Battaglia}, N., {Smith}, K.~M., \& {Sievers}, J.~L.
  2019, \prd, 100, 103532, \dodoi{10.1103/PhysRevD.100.103532}

\bibitem[{{Marcote} {et~al.}(2020){Marcote}, {Nimmo}, {Hessels}, {Tendulkar},
  {Bassa}, {Paragi}, {Keimpema}, {Bhardwaj}, {Karuppusamy}, {Kaspi}, {Law},
  {Michilli}, {Aggarwal}, {Andersen}, {Archibald}, {Bandura}, {Bower}, {Boyle},
  {Brar}, {Burke-Spolaor}, {Butler}, {Cassanelli}, {Chawla}, {Demorest},
  {Dobbs}, {Fonseca}, {Giri}, {Good}, {Gourdji}, {Josephy}, {Kirichenko},
  {Kirsten}, {Landecker}, {Lang}, {Lazio}, {Li}, {Lin}, {Linford}, {Masui},
  {Mena-Parra}, {Naidu}, {Ng}, {Patel}, {Pen}, {Pleunis}, {Rafiei-Ravandi},
  {Rahman}, {Renard}, {Scholz}, {Siegel}, {Smith}, {Stairs}, {Vanderlinde}, \&
  {Zwaniga}}]{marcoteR3}
{Marcote}, B., {Nimmo}, K., {Hessels}, J.~W.~T., {et~al.} 2020, \nat, 577, 190,
  \dodoi{10.1038/s41586-019-1866-z}

\bibitem[{{Marinacci} {et~al.}(2018){Marinacci}, {Vogelsberger}, {Pakmor},
  {Torrey}, {Springel}, {Hernquist}, {Nelson}, {Weinberger}, {Pillepich},
  {Naiman}, \& {Genel}}]{illustriscluster}
{Marinacci}, F., {Vogelsberger}, M., {Pakmor}, R., {et~al.} 2018, \mnras, 480,
  5113, \dodoi{10.1093/mnras/sty2206}

\bibitem[{{Masui} {et~al.}(2015){Masui}, {Lin}, {Sievers}, {Anderson}, {Chang},
  {Chen}, {Ganguly}, {Jarvis}, {Kuo}, {Li}, {Liao}, {McLaughlin}, {Pen},
  {Peterson}, {Roman}, {Timbie}, {Voytek}, \& {Yadav}}]{masui15}
{Masui}, K., {Lin}, H.-H., {Sievers}, J., {et~al.} 2015, \nat, 528, 523,
  \dodoi{10.1038/nature15769}

\bibitem[{{Mazzotta} {et~al.}(2004){Mazzotta}, {Rasia}, {Moscardini}, \&
  {Tormen}}]{Mazzotta2004}
{Mazzotta}, P., {Rasia}, E., {Moscardini}, L., \& {Tormen}, G. 2004, \mnras,
  354, 10, \dodoi{10.1111/j.1365-2966.2004.08167.x}

\bibitem[{{McQuinn}(2016)}]{mcquinn-review}
{McQuinn}, M. 2016, \araa, 54, 313, \dodoi{10.1146/annurev-astro-082214-122355}

\bibitem[{{Mroczkowski} {et~al.}(2019){Mroczkowski}, {Nagai}, {Basu}, {Chluba},
  {Sayers}, {Adam}, {Churazov}, {Crites}, {Di Mascolo}, {Eckert},
  {Macias-Perez}, {Mayet}, {Perotto}, {Pointecouteau}, {Romero}, {Ruppin},
  {Scannapieco}, \& {ZuHone}}]{tsz}
{Mroczkowski}, T., {Nagai}, D., {Basu}, K., {et~al.} 2019, \ssr, 215, 17,
  \dodoi{10.1007/s11214-019-0581-2}

\bibitem[{{Nelson} {et~al.}(2019){Nelson}, {Springel}, {Pillepich},
  {Rodriguez-Gomez}, {Torrey}, {Genel}, {Vogelsberger}, {Pakmor}, {Marinacci},
  {Weinberger}, {Kelley}, {Lovell}, {Diemer}, \& {Hernquist}}]{illustrisTNG}
{Nelson}, D., {Springel}, V., {Pillepich}, A., {et~al.} 2019, Computational
  Astrophysics and Cosmology, 6, 2, \dodoi{10.1186/s40668-019-0028-x}

\bibitem[{{Niu} {et~al.}(2022{\natexlab{a}}){Niu}, {Aggarwal}, {Li}, {Zhang},
  {Chatterjee}, {Tsai}, {Yu}, {Law}, {Burke-Spolaor}, {Cordes}, {Zhang},
  {Ocker}, {Yao}, {Wang}, {Feng}, {Niino}, {Bochenek}, {Cruces}, {Connor},
  {Jiang}, {Dai}, {Luo}, {Li}, {Miao}, {Niu}, {Anna-Thomas}, {Sydnor}, {Stern},
  {Wang}, {Yuan}, {Yue}, {Zhou}, {Yan}, {Zhu}, \& {Zhang}}]{niu190520}
{Niu}, C.~H., {Aggarwal}, K., {Li}, D., {et~al.} 2022{\natexlab{a}}, \nat, 606,
  873, \dodoi{10.1038/s41586-022-04755-5}

\bibitem[{{Niu} {et~al.}(2022{\natexlab{b}}){Niu}, {Aggarwal}, {Li}, {Zhang},
  {Chatterjee}, {Tsai}, {Yu}, {Law}, {Burke-Spolaor}, {Cordes}, {Zhang},
  {Ocker}, {Yao}, {Wang}, {Feng}, {Niino}, {Bochenek}, {Cruces}, {Connor},
  {Jiang}, {Dai}, {Luo}, {Li}, {Miao}, {Niu}, {Anna-Thomas}, {Sydnor}, {Stern},
  {Wang}, {Yuan}, {Yue}, {Zhou}, {Yan}, {Zhu}, \& {Zhang}}]{frb190520}
---. 2022{\natexlab{b}}, \nat, 606, 873, \dodoi{10.1038/s41586-022-04755-5}

\bibitem[{{Ocker} {et~al.}(2022){Ocker}, {Cordes}, {Chatterjee}, \&
  {Gorsuch}}]{ocker2022b}
{Ocker}, S.~K., {Cordes}, J.~M., {Chatterjee}, S., \& {Gorsuch}, M.~R. 2022,
  \apj, 934, 71, \dodoi{10.3847/1538-4357/ac75ba}

\bibitem[{{Ocker} {et~al.}(2023){Ocker}, {Cordes}, {Chatterjee}, {Li}, {Niu},
  {McKee}, {Law}, \& {Anna-Thomas}}]{ocker190520}
{Ocker}, S.~K., {Cordes}, J.~M., {Chatterjee}, S., {et~al.} 2023, \mnras, 519,
  821, \dodoi{10.1093/mnras/stac3547}

\bibitem[{{Oppenheimer} {et~al.}(2021){Oppenheimer}, {Babul}, {Bah{\'e}},
  {Butsky}, \& {McCarthy}}]{Oppenheimer21}
{Oppenheimer}, B.~D., {Babul}, A., {Bah{\'e}}, Y., {Butsky}, I.~S., \&
  {McCarthy}, I.~G. 2021, Universe, 7, 209, \dodoi{10.3390/universe7070209}

\bibitem[{{Pastor-Marazuela} {et~al.}(2021){Pastor-Marazuela}, {Connor}, {van
  Leeuwen}, {Maan}, {ter Veen}, {Bilous}, {Oostrum}, {Petroff}, {Straal},
  {Vohl}, {Attema}, {Boersma}, {Kooistra}, {van der Schuur}, {Sclocco},
  {Smits}, {Adams}, {Adebahr}, {de Blok}, {Coolen}, {Damstra}, {D{\'e}nes},
  {Hess}, {van der Hulst}, {Hut}, {Ivashina}, {Kutkin}, {Loose}, {Lucero},
  {Mika}, {Moss}, {Mulder}, {Norden}, {Oosterloo}, {Orr{\'u}}, {Ruiter}, \&
  {Wijnholds}}]{inesR3}
{Pastor-Marazuela}, I., {Connor}, L., {van Leeuwen}, J., {et~al.} 2021, \nat,
  596, 505, \dodoi{10.1038/s41586-021-03724-8}

\bibitem[{{Petroff} {et~al.}(2019){Petroff}, {Hessels}, \&
  {Lorimer}}]{petroffreview}
{Petroff}, E., {Hessels}, J.~W.~T., \& {Lorimer}, D.~R. 2019, \aapr, 27, 4,
  \dodoi{10.1007/s00159-019-0116-6}

\bibitem[{{Petroff} {et~al.}(2016){Petroff}, {Barr}, {Jameson}, {Keane},
  {Bailes}, {Kramer}, {Morello}, {Tabbara}, \& {van Straten}}]{frbcat}
{Petroff}, E., {Barr}, E.~D., {Jameson}, A., {et~al.} 2016, \pasa, 33, e045,
  \dodoi{10.1017/pasa.2016.35}

\bibitem[{{Piffaretti} {et~al.}(2011){Piffaretti}, {Arnaud}, {Pratt},
  {Pointecouteau}, \& {Melin}}]{mcxc2011}
{Piffaretti}, R., {Arnaud}, M., {Pratt}, G.~W., {Pointecouteau}, E., \&
  {Melin}, J.~B. 2011, \aap, 534, A109, \dodoi{10.1051/0004-6361/201015377}

\bibitem[{{Planck Collaboration} {et~al.}(2014){Planck Collaboration}, {Ade},
  {Aghanim}, {Armitage-Caplan}, {Arnaud}, {Ashdown}, {Atrio-Barandela},
  {Aumont}, {Baccigalupi}, {Banday}, {Barreiro}, {Barrena}, {Bartlett},
  {Battaner}, {Battye}, {Benabed}, {Beno{\^\i}t}, {Benoit-L{\'e}vy}, {Bernard},
  {Bersanelli}, {Bielewicz}, {Bikmaev}, {Blanchard}, {Bobin}, {Bock},
  {B{\"o}hringer}, {Bonaldi}, {Bond}, {Borrill}, {Bouchet}, {Bourdin},
  {Bridges}, {Brown}, {Bucher}, {Burenin}, {Burigana}, {Butler}, {Cardoso},
  {Carvalho}, {Catalano}, {Challinor}, {Chamballu}, {Chary}, {Chiang},
  {Chiang}, {Chon}, {Christensen}, {Church}, {Clements}, {Colombi}, {Colombo},
  {Couchot}, {Coulais}, {Crill}, {Curto}, {Cuttaia}, {Da Silva}, {Dahle},
  {Danese}, {Davies}, {Davis}, {de Bernardis}, {de Rosa}, {de Zotti},
  {Delabrouille}, {Delouis}, {D{\'e}mocl{\`e}s}, {D{\'e}sert}, {Dickinson},
  {Diego}, {Dolag}, {Dole}, {Donzelli}, {Dor{\'e}}, {Douspis}, {Dupac},
  {Efstathiou}, {En{\ss}lin}, {Eriksen}, {Finelli}, {Flores-Cacho}, {Forni},
  {Frailis}, {Franceschi}, {Fromenteau}, {Galeotta}, {Ganga},
  {G{\'e}nova-Santos}, {Giard}, {Giardino}, {Giraud-H{\'e}raud},
  {Gonz{\'a}lez-Nuevo}, {G{\'o}rski}, {Gratton}, {Gregorio}, {Gruppuso},
  {Hansen}, {Hanson}, {Harrison}, {Henrot-Versill{\'e}},
  {Hern{\'a}ndez-Monteagudo}, {Herranz}, {Hildebrandt}, {Hivon}, {Hobson},
  {Holmes}, {Hornstrup}, {Hovest}, {Huffenberger}, {Hurier}, {Jaffe}, {Jaffe},
  {Jones}, {Juvela}, {Keih{\"a}nen}, {Keskitalo}, {Khamitov}, {Kisner},
  {Kneissl}, {Knoche}, {Knox}, {Kunz}, {Kurki-Suonio}, {Lagache},
  {L{\"a}hteenm{\"a}ki}, {Lamarre}, {Lasenby}, {Laureijs}, {Lawrence}, {Leahy},
  {Leonardi}, {Le{\'o}n-Tavares}, {Lesgourgues}, {Liddle}, {Liguori}, {Lilje},
  {Linden-V{\o}rnle}, {L{\'o}pez-Caniego}, {Lubin}, {Mac{\'\i}as-P{\'e}rez},
  {Maffei}, {Maino}, {Mandolesi}, {Marcos-Caballero}, {Maris}, {Marshall},
  {Martin}, {Mart{\'\i}nez-Gonz{\'a}lez}, {Masi}, {Matarrese}, {Matthai},
  {Mazzotta}, {Meinhold}, {Melchiorri}, {Melin}, {Mendes}, {Mennella},
  {Migliaccio}, {Mitra}, {Miville-Desch{\^e}nes}, {Moneti}, {Montier},
  {Morgante}, {Mortlock}, {Moss}, {Munshi}, {Naselsky}, {Nati}, {Natoli},
  {Netterfield}, {N{\o}rgaard-Nielsen}, {Noviello}, {Novikov}, {Novikov},
  {Osborne}, {Oxborrow}, {Paci}, {Pagano}, {Pajot}, {Paoletti}, {Partridge},
  {Pasian}, {Patanchon}, {Perdereau}, {Perotto}, {Perrotta}, {Piacentini},
  {Piat}, {Pierpaoli}, {Pietrobon}, {Plaszczynski}, {Pointecouteau}, {Polenta},
  {Ponthieu}, {Popa}, {Poutanen}, {Pratt}, {Pr{\'e}zeau}, {Prunet}, {Puget},
  {Rachen}, {Rebolo}, {Reinecke}, {Remazeilles}, {Renault}, {Ricciardi},
  {Riller}, {Ristorcelli}, {Rocha}, {Roman}, {Rosset}, {Roudier},
  {Rowan-Robinson}, {Rubi{\~n}o-Mart{\'\i}n}, {Rusholme}, {Sandri}, {Santos},
  {Savini}, {Scott}, {Seiffert}, {Shellard}, {Spencer}, {Starck}, {Stolyarov},
  {Stompor}, {Sudiwala}, {Sunyaev}, {Sureau}, {Sutton}, {Suur-Uski}, {Sygnet},
  {Tauber}, {Tavagnacco}, {Terenzi}, {Toffolatti}, {Tomasi}, {Tristram},
  {Tucci}, {Tuovinen}, {T{\"u}rler}, {Umana}, {Valenziano}, {Valiviita}, {Van
  Tent}, {Vielva}, {Villa}, {Vittorio}, {Wade}, {Wandelt}, {Weller}, {White},
  {White}, {Yvon}, {Zacchei}, \& {Zonca}}]{planck14}
{Planck Collaboration}, {Ade}, P.~A.~R., {Aghanim}, N., {et~al.} 2014, \aap,
  571, A20, \dodoi{10.1051/0004-6361/201321521}

\bibitem[{{Planck Collaboration} {et~al.}(2016{\natexlab{a}}){Planck
  Collaboration}, {Ade}, {Aghanim}, {Arnaud}, {Ashdown}, {Aumont},
  {Baccigalupi}, {Banday}, {Barreiro}, {Barrena}, {Bartlett}, {Bartolo},
  {Battaner}, {Battye}, {Benabed}, {Beno{\^\i}t}, {Benoit-L{\'e}vy}, {Bernard},
  {Bersanelli}, {Bielewicz}, {Bikmaev}, {B{\"o}hringer}, {Bonaldi}, {Bonavera},
  {Bond}, {Borrill}, {Bouchet}, {Bucher}, {Burenin}, {Burigana}, {Butler},
  {Calabrese}, {Cardoso}, {Carvalho}, {Catalano}, {Challinor}, {Chamballu},
  {Chary}, {Chiang}, {Chon}, {Christensen}, {Clements}, {Colombi}, {Colombo},
  {Combet}, {Comis}, {Couchot}, {Coulais}, {Crill}, {Curto}, {Cuttaia},
  {Dahle}, {Danese}, {Davies}, {Davis}, {de Bernardis}, {de Rosa}, {de Zotti},
  {Delabrouille}, {D{\'e}sert}, {Dickinson}, {Diego}, {Dolag}, {Dole},
  {Donzelli}, {Dor{\'e}}, {Douspis}, {Ducout}, {Dupac}, {Efstathiou},
  {Eisenhardt}, {Elsner}, {En{\ss}lin}, {Eriksen}, {Falgarone}, {Fergusson},
  {Feroz}, {Ferragamo}, {Finelli}, {Forni}, {Frailis}, {Fraisse}, {Franceschi},
  {Frejsel}, {Galeotta}, {Galli}, {Ganga}, {G{\'e}nova-Santos}, {Giard},
  {Giraud-H{\'e}raud}, {Gjerl{\o}w}, {Gonz{\'a}lez-Nuevo}, {G{\'o}rski},
  {Grainge}, {Gratton}, {Gregorio}, {Gruppuso}, {Gudmundsson}, {Hansen},
  {Hanson}, {Harrison}, {Hempel}, {Henrot-Versill{\'e}},
  {Hern{\'a}ndez-Monteagudo}, {Herranz}, {Hildebrandt}, {Hivon}, {Hobson},
  {Holmes}, {Hornstrup}, {Hovest}, {Huffenberger}, {Hurier}, {Jaffe}, {Jaffe},
  {Jin}, {Jones}, {Juvela}, {Keih{\"a}nen}, {Keskitalo}, {Khamitov}, {Kisner},
  {Kneissl}, {Knoche}, {Kunz}, {Kurki-Suonio}, {Lagache}, {Lamarre}, {Lasenby},
  {Lattanzi}, {Lawrence}, {Leonardi}, {Lesgourgues}, {Levrier}, {Liguori},
  {Lilje}, {Linden-V{\o}rnle}, {L{\'o}pez-Caniego}, {Lubin},
  {Mac{\'\i}as-P{\'e}rez}, {Maggio}, {Maino}, {Mak}, {Mandolesi}, {Mangilli},
  {Martin}, {Mart{\'\i}nez-Gonz{\'a}lez}, {Masi}, {Matarrese}, {Mazzotta},
  {McGehee}, {Mei}, {Melchiorri}, {Melin}, {Mendes}, {Mennella}, {Migliaccio},
  {Mitra}, {Miville-Desch{\^e}nes}, {Moneti}, {Montier}, {Morgante},
  {Mortlock}, {Moss}, {Munshi}, {Murphy}, {Naselsky}, {Nastasi}, {Nati},
  {Natoli}, {Netterfield}, {N{\o}rgaard-Nielsen}, {Noviello}, {Novikov},
  {Novikov}, {Olamaie}, {Oxborrow}, {Paci}, {Pagano}, {Pajot}, {Paoletti},
  {Pasian}, {Patanchon}, {Pearson}, {Perdereau}, {Perotto}, {Perrott},
  {Perrotta}, {Pettorino}, {Piacentini}, {Piat}, {Pierpaoli}, {Pietrobon},
  {Plaszczynski}, {Pointecouteau}, {Polenta}, {Pratt}, {Pr{\'e}zeau}, {Prunet},
  {Puget}, {Rachen}, {Reach}, {Rebolo}, {Reinecke}, {Remazeilles}, {Renault},
  {Renzi}, {Ristorcelli}, {Rocha}, {Rosset}, {Rossetti}, {Roudier}, {Rozo},
  {Rubi{\~n}o-Mart{\'\i}n}, {Rumsey}, {Rusholme}, {Rykoff}, {Sandri}, {Santos},
  {Saunders}, {Savelainen}, {Savini}, {Schammel}, {Scott}, {Seiffert},
  {Shellard}, {Shimwell}, {Spencer}, {Stanford}, {Stern}, {Stolyarov},
  {Stompor}, {Streblyanska}, {Sudiwala}, {Sunyaev}, {Sutton}, {Suur-Uski},
  {Sygnet}, {Tauber}, {Terenzi}, {Toffolatti}, {Tomasi}, {Tramonte},
  {Tristram}, {Tucci}, {Tuovinen}, {Umana}, {Valenziano}, {Valiviita}, {Van
  Tent}, {Vielva}, {Villa}, {Wade}, {Wandelt}, {Wehus}, {White}, {Wright},
  {Yvon}, {Zacchei}, \& {Zonca}}]{PSZ2}
---. 2016{\natexlab{a}}, \aap, 594, A27, \dodoi{10.1051/0004-6361/201525823}

\bibitem[{{Planck Collaboration} {et~al.}(2016{\natexlab{b}}){Planck
  Collaboration}, {Aghanim}, {Arnaud}, {Ashdown}, {Aumont}, {Baccigalupi},
  {Banday}, {Barreiro}, {Bartlett}, {Bartolo}, {Battaner}, {Battye}, {Benabed},
  {Beno{\^\i}t}, {Benoit-L{\'e}vy}, {Bernard}, {Bersanelli}, {Bielewicz},
  {Bock}, {Bonaldi}, {Bonavera}, {Bond}, {Borrill}, {Bouchet}, {Burigana},
  {Butler}, {Calabrese}, {Cardoso}, {Catalano}, {Challinor}, {Chiang},
  {Christensen}, {Churazov}, {Clements}, {Colombo}, {Combet}, {Comis},
  {Coulais}, {Crill}, {Curto}, {Cuttaia}, {Danese}, {Davies}, {Davis}, {de
  Bernardis}, {de Rosa}, {de Zotti}, {Delabrouille}, {D{\'e}sert}, {Dickinson},
  {Diego}, {Dolag}, {Dole}, {Donzelli}, {Dor{\'e}}, {Douspis}, {Ducout},
  {Dupac}, {Efstathiou}, {Elsner}, {En{\ss}lin}, {Eriksen}, {Fergusson},
  {Finelli}, {Forni}, {Frailis}, {Fraisse}, {Franceschi}, {Frejsel},
  {Galeotta}, {Galli}, {Ganga}, {G{\'e}nova-Santos}, {Giard},
  {Gonz{\'a}lez-Nuevo}, {G{\'o}rski}, {Gregorio}, {Gruppuso}, {Gudmundsson},
  {Hansen}, {Harrison}, {Henrot-Versill{\'e}}, {Hern{\'a}ndez-Monteagudo},
  {Herranz}, {Hildebrandt}, {Hivon}, {Holmes}, {Hornstrup}, {Huffenberger},
  {Hurier}, {Jaffe}, {Jones}, {Juvela}, {Keih{\"a}nen}, {Keskitalo}, {Kneissl},
  {Knoche}, {Kunz}, {Kurki-Suonio}, {Lacasa}, {Lagache}, {L{\"a}hteenm{\"a}ki},
  {Lamarre}, {Lasenby}, {Lattanzi}, {Leonardi}, {Lesgourgues}, {Levrier},
  {Liguori}, {Lilje}, {Linden-V{\o}rnle}, {L{\'o}pez-Caniego},
  {Mac{\'\i}as-P{\'e}rez}, {Maffei}, {Maggio}, {Maino}, {Mandolesi},
  {Mangilli}, {Maris}, {Martin}, {Mart{\'\i}nez-Gonz{\'a}lez}, {Masi},
  {Matarrese}, {Melchiorri}, {Melin}, {Migliaccio}, {Miville-Desch{\^e}nes},
  {Moneti}, {Montier}, {Morgante}, {Mortlock}, {Munshi}, {Murphy}, {Naselsky},
  {Nati}, {Natoli}, {Noviello}, {Novikov}, {Novikov}, {Paci}, {Pagano},
  {Pajot}, {Paoletti}, {Pasian}, {Patanchon}, {Perdereau}, {Perotto},
  {Pettorino}, {Piacentini}, {Piat}, {Pierpaoli}, {Pietrobon}, {Plaszczynski},
  {Pointecouteau}, {Polenta}, {Ponthieu}, {Pratt}, {Prunet}, {Puget}, {Rachen},
  {Reinecke}, {Remazeilles}, {Renault}, {Renzi}, {Ristorcelli}, {Rocha},
  {Rossetti}, {Roudier}, {Rubi{\~n}o-Mart{\'\i}n}, {Rusholme}, {Sandri},
  {Santos}, {Sauv{\'e}}, {Savelainen}, {Savini}, {Scott}, {Spencer},
  {Stolyarov}, {Stompor}, {Sunyaev}, {Sutton}, {Suur-Uski}, {Sygnet}, {Tauber},
  {Terenzi}, {Toffolatti}, {Tomasi}, {Tramonte}, {Tristram}, {Tucci},
  {Tuovinen}, {Valenziano}, {Valiviita}, {Van Tent}, {Vielva}, {Villa}, {Wade},
  {Wandelt}, {Wehus}, {Yvon}, {Zacchei}, \& {Zonca}}]{Planck2016_ymap}
{Planck Collaboration}, {Aghanim}, N., {Arnaud}, M., {et~al.}
  2016{\natexlab{b}}, \aap, 594, A22,
  \dodoi{10.1051/0004-6361/20152582610.48550/arXiv.1502.01596}

\bibitem[{{Planck Collaboration} {et~al.}(2016{\natexlab{c}}){Planck
  Collaboration}, {Ade}, {Aghanim}, {Arnaud}, {Ashdown}, {Aumont},
  {Baccigalupi}, {Banday}, {Barreiro}, {Bartolo}, {Battaner}, {Benabed},
  {Benoit-L{\'e}vy}, {Bernard}, {Bersanelli}, {Bielewicz}, {Bonaldi},
  {Bonavera}, {Bond}, {Borrill}, {Bouchet}, {Burigana}, {Butler}, {Calabrese},
  {Cardoso}, {Catalano}, {Chamballu}, {Chiang}, {Christensen}, {Churazov},
  {Clements}, {Colombo}, {Combet}, {Comis}, {Couchot}, {Coulais}, {Crill},
  {Curto}, {Cuttaia}, {Danese}, {Davies}, {Davis}, {de Bernardis}, {de Rosa},
  {de Zotti}, {Delabrouille}, {Dickinson}, {Diego}, {Dolag}, {Dole},
  {Donzelli}, {Dor{\'e}}, {Douspis}, {Ducout}, {Dupac}, {Efstathiou}, {Elsner},
  {En{\ss}lin}, {Eriksen}, {Finelli}, {Forni}, {Frailis}, {Fraisse},
  {Franceschi}, {Galeotta}, {Galli}, {Ganga}, {Giard}, {Giraud-H{\'e}raud},
  {Gjerl{\o}w}, {Gonz{\'a}lez-Nuevo}, {G{\'o}rski}, {Gregorio}, {Gruppuso},
  {Gudmundsson}, {Hansen}, {Harrison}, {Helou}, {Hern{\'a}ndez-Monteagudo},
  {Herranz}, {Hildebrandt}, {Hivon}, {Hobson}, {Hornstrup}, {Hovest},
  {Huffenberger}, {Hurier}, {Jaffe}, {Jaffe}, {Jones}, {Keih{\"a}nen},
  {Keskitalo}, {Kisner}, {Kneissl}, {Knoche}, {Kunz}, {Kurki-Suonio},
  {Lagache}, {Lamarre}, {Lasenby}, {Lattanzi}, {Lawrence}, {Leonardi},
  {Levrier}, {Liguori}, {Lilje}, {Linden-V{\o}rnle}, {L{\'o}pez-Caniego},
  {Lubin}, {Mac{\'\i}as-P{\'e}rez}, {Maffei}, {Maggio}, {Maino}, {Mandolesi},
  {Mangilli}, {Marcos-Caballero}, {Maris}, {Martin},
  {Mart{\'\i}nez-Gonz{\'a}lez}, {Masi}, {Matarrese}, {Mazzotta}, {Meinhold},
  {Melchiorri}, {Mennella}, {Migliaccio}, {Mitra}, {Miville-Desch{\^e}nes},
  {Moneti}, {Montier}, {Morgante}, {Mortlock}, {Munshi}, {Murphy}, {Naselsky},
  {Nati}, {Natoli}, {Noviello}, {Novikov}, {Novikov}, {Oppermann}, {Oxborrow},
  {Pagano}, {Pajot}, {Paoletti}, {Pasian}, {Pearson}, {Perdereau}, {Perotto},
  {Pettorino}, {Piacentini}, {Piat}, {Pierpaoli}, {Plaszczynski},
  {Pointecouteau}, {Polenta}, {Ponthieu}, {Pratt}, {Prunet}, {Puget}, {Rachen},
  {Reinecke}, {Remazeilles}, {Renault}, {Renzi}, {Ristorcelli}, {Rocha},
  {Rosset}, {Rossetti}, {Roudier}, {Rubi{\~n}o-Mart{\'\i}n}, {Rusholme},
  {Sandri}, {Santos}, {Savelainen}, {Savini}, {Schaefer}, {Scott}, {Soler},
  {Stolyarov}, {Stompor}, {Sudiwala}, {Sunyaev}, {Sutton}, {Suur-Uski},
  {Sygnet}, {Tauber}, {Terenzi}, {Toffolatti}, {Tomasi}, {Tristram}, {Tucci},
  {Umana}, {Valenziano}, {Valiviita}, {Van Tent}, {Vielva}, {Villa}, {Wade},
  {Wandelt}, {Wehus}, {Weller}, {Yvon}, {Zacchei}, \& {Zonca}}]{PSZvirgo}
{Planck Collaboration}, {Ade}, P.~A.~R., {Aghanim}, N., {et~al.}
  2016{\natexlab{c}}, \aap, 596, A101, \dodoi{10.1051/0004-6361/201527743}

\bibitem[{{Pratt} {et~al.}(2007){Pratt}, {B{\"o}hringer}, {Croston}, {Arnaud},
  {Borgani}, {Finoguenov}, \& {Temple}}]{Pratt2007}
{Pratt}, G.~W., {B{\"o}hringer}, H., {Croston}, J.~H., {et~al.} 2007, \aap,
  461, 71, \dodoi{10.1051/0004-6361:20065676}

\bibitem[{{Pratt} {et~al.}(2009{\natexlab{a}}){Pratt}, {Croston}, {Arnaud}, \&
  {B{\"o}hringer}}]{Pratt}
{Pratt}, G.~W., {Croston}, J.~H., {Arnaud}, M., \& {B{\"o}hringer}, H.
  2009{\natexlab{a}}, \aap, 498, 361, \dodoi{10.1051/0004-6361/200810994}

\bibitem[{{Pratt} {et~al.}(2009{\natexlab{b}}){Pratt}, {Croston}, {Arnaud}, \&
  {B{\"o}hringer}}]{pratt2009}
---. 2009{\natexlab{b}}, \aap, 498, 361, \dodoi{10.1051/0004-6361/200810994}

\bibitem[{{Prochaska} \& {Zheng}(2019)}]{prochaska2019b}
{Prochaska}, J.~X., \& {Zheng}, Y. 2019, \mnras, 485, 648,
  \dodoi{10.1093/mnras/stz261}

\bibitem[{{Ravi}(2019)}]{ravi2019}
{Ravi}, V. 2019, \apj, 872, 88, \dodoi{10.3847/1538-4357/aafb30}

\bibitem[{{Ravi} {et~al.}(2019){Ravi}, {Catha}, {D'Addario}, {Djorgovski},
  {Hallinan}, {Hobbs}, {Kocz}, {Kulkarni}, {Shi}, {Vedantham}, {Weinreb}, \&
  {Woody}}]{ravi19frb}
{Ravi}, V., {Catha}, M., {D'Addario}, L., {et~al.} 2019, \nat, 572, 352,
  \dodoi{10.1038/s41586-019-1389-7}

\bibitem[{{Ravi} {et~al.}(2022){Ravi}, {Catha}, {Chen}, {Connor}, {Faber},
  {Lamb}, {Hallinan}, {Harnach}, {Hellbourg}, {Hobbs}, {Hodge}, {Hodges},
  {Law}, {Rasmussen}, {Sharma}, {Sherman}, {Shi}, {Simard}, {Squillace},
  {Weinreb}, {Woody}, {Yadlapalli}, {Ahumada}, {Dong}, {Fremling}, {Huang},
  {Karambelkar}, \& {Miller}}]{rcc+22}
{Ravi}, V., {Catha}, M., {Chen}, G., {et~al.} 2022, arXiv e-prints,
  arXiv:2211.09049, \dodoi{10.48550/arXiv.2211.09049}

\bibitem[{{Ravi} {et~al.}(2023){Ravi}, {Catha}, {Chen}, {Connor}, {Cordes},
  {Faber}, {Lamb}, {Hallinan}, {Harnach}, {Hellbourg}, {Hobbs}, {Hodge},
  {Hodges}, {Law}, {Rasmussen}, {Sharma}, {Sherman}, {Shi}, {Simard},
  {Somalwar}, {Squillace}, {Weinreb}, {Woody}, \& {Yadlapalli}}]{ravi-mark2023}
---. 2023, arXiv e-prints, arXiv:2301.01000.
\newblock \doarXiv{2301.01000}

\bibitem[{{Sarazin}(1986)}]{cluster-xray}
{Sarazin}, C.~L. 1986, Reviews of Modern Physics, 58, 1,
  \dodoi{10.1103/RevModPhys.58.1}

\bibitem[{{Sayers} {et~al.}(2016){Sayers}, {Golwala}, {Mantz}, {Merten},
  {Molnar}, {Naka}, {Pailet}, {Pierpaoli}, {Siegel}, \& {Wolman}}]{sayers16}
{Sayers}, J., {Golwala}, S.~R., {Mantz}, A.~B., {et~al.} 2016, \apj, 832, 26,
  \dodoi{10.3847/0004-637X/832/1/26}

\bibitem[{{Seta} {et~al.}(2021){Seta}, {Rodrigues}, {Federrath}, \&
  {Hales}}]{seta21}
{Seta}, A., {Rodrigues}, L. F.~S., {Federrath}, C., \& {Hales}, C.~A. 2021,
  \apj, 907, 2, \dodoi{10.3847/1538-4357/abd2bb}

\bibitem[{{Simard} \& {Ravi}(2021)}]{simard21}
{Simard}, D., \& {Ravi}, V. 2021, arXiv e-prints, arXiv:2107.11334.
\newblock \doarXiv{2107.11334}

\bibitem[{{Tendulkar} {et~al.}(2017){Tendulkar}, {Bassa}, {Cordes}, {Bower},
  {Law}, {Chatterjee}, {Adams}, {Bogdanov}, {Burke-Spolaor}, {Butler},
  {Demorest}, {Hessels}, {Kaspi}, {Lazio}, {Maddox}, {Marcote}, {McLaughlin},
  {Paragi}, {Ransom}, {Scholz}, {Seymour}, {Spitler}, {van Langevelde}, \&
  {Wharton}}]{frb121102}
{Tendulkar}, S.~P., {Bassa}, C.~G., {Cordes}, J.~M., {et~al.} 2017, \apjl, 834,
  L7, \dodoi{10.3847/2041-8213/834/2/L7}

\bibitem[{{Tumlinson} {et~al.}(2017){Tumlinson}, {Peeples}, \&
  {Werk}}]{Tumlinson17}
{Tumlinson}, J., {Peeples}, M.~S., \& {Werk}, J.~K. 2017, \araa, 55, 389,
  \dodoi{10.1146/annurev-astro-091916-055240}

\bibitem[{{van Weeren} {et~al.}(2019){van Weeren}, {de Gasperin}, {Akamatsu},
  {Br{\"u}ggen}, {Feretti}, {Kang}, {Stroe}, \& {Zandanel}}]{RMweer}
{van Weeren}, R.~J., {de Gasperin}, F., {Akamatsu}, H., {et~al.} 2019, \ssr,
  215, 16, \dodoi{10.1007/s11214-019-0584-z}

\bibitem[{{Vikhlinin} {et~al.}(2005){Vikhlinin}, {Markevitch}, {Murray},
  {Jones}, {Forman}, \& {Van Speybroeck}}]{Vikhlinin2005}
{Vikhlinin}, A., {Markevitch}, M., {Murray}, S.~S., {et~al.} 2005, \apj, 628,
  655, \dodoi{10.1086/431142}

\bibitem[{{Voit}(2005)}]{cluster-reviewx}
{Voit}, G.~M. 2005, Reviews of Modern Physics, 77, 207,
  \dodoi{10.1103/RevModPhys.77.207}

\bibitem[{{White} {et~al.}(2011){White}, {Daw}, \& {Dhillon}}]{gwgc}
{White}, D.~J., {Daw}, E.~J., \& {Dhillon}, V.~S. 2011, Classical and Quantum
  Gravity, 28, 085016, \dodoi{10.1088/0264-9381/28/8/085016}

\bibitem[{{Xu} {et~al.}(2022){Xu}, {Ramos-Ceja}, {Pacaud}, {Reiprich}, \&
  {Erben}}]{weiwei-rosat}
{Xu}, W., {Ramos-Ceja}, M.~E., {Pacaud}, F., {Reiprich}, T.~H., \& {Erben}, T.
  2022, \aap, 658, A59, \dodoi{10.1051/0004-6361/202140908}

\bibitem[{{Yang} {et~al.}(2021){Yang}, {Xu}, {He}, {Gu}, {Katsianis}, {Meng},
  {Shi}, {Zou}, {Zhang}, {Liu}, {Wang}, {Dong}, {Lu}, {Li}, {Chen}, {Wang},
  {Mo}, {Fu}, {Guo}, {Leauthaud}, {Luo}, {Zhang}, \& {Zu}}]{desicluster}
{Yang}, X., {Xu}, H., {He}, M., {et~al.} 2021, \apj, 909, 143,
  \dodoi{10.3847/1538-4357/abddb2}

\bibitem[{{Yao} {et~al.}(2017){Yao}, {Manchester}, \& {Wang}}]{ymw16}
{Yao}, J.~M., {Manchester}, R.~N., \& {Wang}, N. 2017, \apj, 835, 29,
  \dodoi{10.3847/1538-4357/835/1/29}

\bibitem[{{Zhang} {et~al.}(2021){Zhang}, {Yan}, {Li}, {Zhang}, \&
  {Wang}}]{zhangDM2021}
{Zhang}, Z.~J., {Yan}, K., {Li}, C.~M., {Zhang}, G.~Q., \& {Wang}, F.~Y. 2021,
  \apj, 906, 49, \dodoi{10.3847/1538-4357/abceb9}

\end{thebibliography}
\bibliographystyle{aasjournal}

\end{document}